\documentclass[sn-aps,iicol]{sn-jnl}

\pdfoutput=1

\usepackage{graphicx}%
\usepackage{multirow}%
\usepackage{amsmath,amssymb,amsfonts}%
\usepackage{amsthm}%
\usepackage{mathrsfs}%
\usepackage[title]{appendix}%
\usepackage{xcolor}%
\usepackage{textcomp}%
\usepackage{manyfoot}%
\usepackage{booktabs}%
\usepackage{algorithm}%
\usepackage{algorithmicx}%
\usepackage{algpseudocode}%
\usepackage{listings}%

\usepackage[utf8]{inputenc}
\usepackage{amsmath}
\usepackage{color}
\usepackage{graphicx}
\usepackage[normalem]{ulem}
\usepackage{url}
\makeatletter
\g@addto@macro{\UrlBreaks}{\UrlOrds}
\makeatother

\begin{document}

\title[A C++ program for estimating detector sensitivities to long-lived particles: Displaced Decay Counter]{A C++ program for estimating detector sensitivities to long-lived particles: Displaced Decay Counter}

\author[a]{\fnm{Florian} \sur{Domingo}}\email{domingo@physik.uni-bonn.de}

\author[a]{\fnm{Julian} \sur{G\"unther}}\email{guenther@physik.uni-bonn.de}

\author[b]{\fnm{Jong Soo} \sur{Kim}}\email{jongsoo.kim@tu-dortmund.de}

\author[c,d]{\fnm{Zeren Simon} \sur{Wang}}\email{wzs@mx.nthu.edu.tw}

\affil[a]{Bethe Center for Theoretical Physics and Physikalisches, Institut der Universit\"at Bonn, Nu\ss allee 12, 53115 Bonn, Germany}

\affil[b]{National Institute for Theoretical Physics and School of Physics, University of the Witwatersrand, Johannesburg, South Africa}

\affil[c]{Department of Physics, National Tsing Hua University, Hsinchu 300, Taiwan}

\affil[d]{Center for Theory and Computation, National Tsing Hua University, Hsinchu 300, Taiwan}

\abstract{A series of far-detector programs have been proposed for operation at various interaction points of the Large Hadron Collider during the upcoming runs.
	Investigating the potential and complementarity of these experiments for new-physics searches goes through the estimation of their sensitivity to specific long-lived particle models.
	Here, we present an integrated numerical tool written in the C++ language and called \texttt{Displaced Decay Counter}, which we have created to this end and which can be used in association with
	\texttt{MadGraph5}, \texttt{Pythia8}, or any other state-of-the-art Monte-Carlo collider simulation tool. 
	Several far-detector models have been implemented within the program, accounting for the geometry and integrated luminosity of projected detectors. 
	Additional or more accurate designs can be easily constructed through a dedicated interface.
	The functionality of this tool is exemplified through the discussion of several benchmark scenarios, which we consider for the validation of the implemented detector models.}




\maketitle

\section{Introduction}\label{sec:intro}
A decade of operation at the Large Hadron Collider (LHC) has left a picture of particle physics at energies below the TeV scale which seems well compatible with a Standard Model (SM) interpretation.
In particular, no conclusive evidence has emerged to corroborate the existence of heavy color- or electroweakly-charged resonances leaving massive missing-energy signatures, as expected in various models of new-physics where a $Z_2$ symmetry distinguishes between the standard and exotic sectors --- such as $R$-parity in supersymmetry (SUSY)-inspired models or $T$-parity in constructions involving a new strongly-coupled interaction.
While the ensuing constraints leave substantial parameter space (possibly inaccessible to the LHC) in the scenarios of the previous type, it is desirable to exploit as much information as TeV-scale colliders can provide us with for the searches of physics signatures beyond the SM.
Long-lived particles (LLPs) taking mass at the electroweak scale form a well-motivated alternative to the prompt-decay picture, more challenging to test, and may thus have been overlooked in traditional collider search strategies.
A long lifetime in an exotic particle may appear as the result of the protection by an approximate symmetry (e.g.~of the `matter-parity' type), compressed phase space, or a portal to the SM involving very massive mediators, for instance.
Consequently, LLPs are widely predicted in various new-physics models~\cite{Curtin:2018mvb}.
We refer the reader to Refs.~\cite{Lee:2018pag,Beacham:2019nyx,Alimena:2019zri,Agrawal:2021dbo,Knapen:2022afb} for some recent reviews on LLP searches.

Several initiatives have emerged in the last few years, both from the experimental and phenomenological communities, to promote the search for LLPs in LHC-based experiments.
First, a number of LLP signatures can be tested at existing detectors, such as ATLAS and CMS (see e.g.~Refs.~\cite{ATLAS:2022atq,ATLAS:2022zhj,ATLAS:2022bll,CMS:2022qej} for some recent searches in this direction), and allow to constrain displaced leptons~\cite{CMS:2021lcp}, displaced vertices with missing transverse momentum~\cite{ATLAS:2017tny}, leptons~\cite{ATLAS:2020xyo}, or jets~\cite{ATLAS:2023oti}, or heavy charged tracks~\cite{ATLAS:2019gqq}, to restrict to a few examples.
It is then possible to recast such limits and make them available for testing arbitrary models~\cite{Desai:2021jsa,Araz:2021akd,Alguero:2021dig,LLPrecastingRepo,delphes_pr}.
However, such searches usually focus on macroscopic, but short lifetimes, so as to account for the proximity of the detector with the interaction point.
Far detectors offer a complementary search strategy, as they would be sensitive to exotic particles decaying further away.
In addition, the SM background is anticipated to be very restricted, or even negligible.
A collection of designs for such far detectors has flourished in recent years, some of which have been approved and are scheduled for operation in the current LHC Run3.
While the detail of the efficiencies is in general unavailable, it is still enlightening to examine the exclusion potential of the projected detectors and their complementarity in constraining various types of new physics models.
Simulations of this type have been carried out in numerous analyses, using a more or less detailed modelization of the detectors to estimate their acceptance in purely geometrical terms (see e.g.~Refs.~\cite{FASER:2018eoc,Helo:2018qej,Dercks:2018eua,Dercks:2018wum,Hirsch:2020klk} for a few existing works).
Specific studies on optimization of the geometry of such far detectors can also be found in e.g.~Refs.~\cite{
	Gorordo:2022rro,Plows:2022gxc}.
Moreover, recasting works on the sensitivities of these experiments can be found in e.g.~Refs.~\cite{
	Beltran:2023nli,Fernandez-Martinez:2023phj,Dreiner:2023gir}.
The purpose of this paper is to present an integrated tool allowing to assess the sensitivity of multiple far detectors in an efficient and systematic fashion: \texttt{Displaced Decay Counter (DDC)}.

Several tools have been presented recently that accomplish somewhat comparable functions to those of \texttt{DDC}.
First, \texttt{MadDump}~\cite{Buonocore:2018xjk} is a plugin for \texttt{MadGraph5\_aMC@NLO}~\cite{Alwall:2011uj,Alwall:2014hca} and dedicated to the computation of signal-event rates of LLPs including their decays, scattering off SM particles, and their far detection, at beam-dump experiments.
It takes into account the LLP kinematics as well as the geometry of the experiments.
It can also be employed for the simulation of non-beam-dump processes, in principle.
Next, \texttt{FORESEE}~\cite{Kling:2021fwx} is a \texttt{Python}-based code that calculates the sensitivity reach of far detectors at the LHC and FCC-hh.
This package computes the kinematical distributions of LLPs with the help of tables of kinematical distributions of various SM particles, allowing to derive the signal-event rates at the LHC far detectors.
The event records are generated in the \texttt{HepMC} format.
Although currently only a limited number of models have been implemented, the code can be extended to further LLP environments.
A cut functionality is available but the efficiencies of decay products are not taken into account.
Then, the authors of Ref.~\cite{Jerhot:2022chi} implemented the tool \texttt{ALPINIST}, which estimates the signal-event rates at extracted-beam experiments, for axion-like particles (ALPs) with generic couplings.
Making use of \texttt{Mathematica}, \texttt{ROOT}, and \texttt{Python}, the code reconstructs the final-state particles of the ALP inside a detector and performs the computation with fully reconstructed final states.
Furthermore, the new package \texttt{SensCalc} was presented in Ref.~\cite{Ovchynnikov:2023cry}.
This code employs a pipeline of \texttt{Mathematica} notebooks in order to calculate the detector acceptance, sequentially deriving the LLP distributions, computing the signal-event rates, and plotting.
The LLP distributions are either pre-defined or can be computed by running Monte-Carlo (MC) simulation.
Portal-physics models --- heavy neutral leptons, dark scalars, dark photons, and axion-like particles  --- have already been considered in \texttt{SensCalc}, and it could be extended to further LLP models.
Again \texttt{SensCalc} does not reconstruct the final states.
Finally, the authors of Ref.~\cite{Curtin:2023skh} very recently published a Python module \texttt{FastSim} that allows efficient signal-event studies of LLP decays in various far detectors.
The tool includes idealized reconstruction criteria for displaced vertices, such as the number of observable charged particles from the LLP decays that should pass a certain momentum threshold and also intersect at least a certain number of detector planes.

The main functionality of the tool \texttt{DDC} under discussion is to compute the decay probability of an LLP within a pre-defined detector volume.
A collection of detector models based on LHC designs is already pre-encoded and an integrated editor allows for further construction, possibly applying to other types of colliders.
MC-simulation events in the \texttt{LHEF}~\cite{Alwall:2006yp} or \texttt{HepMC}~\cite{Dobbs:2001ck}\footnote{The code is only compatible with \texttt{HepMC2} at the current stage.} format can be used as input, as well as \texttt{CMND} input files for internal generation through a \texttt{Pythia8}~\cite{Sjostrand:2014zea} interface.
LLPs are identified for testing via their PDG code.
Compared to the existing codes discussed above, our package \texttt{DDC} completely relies on linked tools such as \texttt{MadGraph5} and \texttt{Pythia8} for performing MC simulation and obtaining the LLP distributions, without any attempt at exploiting pre-generated tabulated densities.
Similarly to \texttt{SensCalc}, \texttt{DDC} is not restricted to a certain type of experiments, but should be easily applicable to a wide range of experimental facilities.
Furthermore, \texttt{DDC} achieves a high degree of versatility: it is indeed compatible with several MC-simulation tools and can be used for a wide range of theoretical models or interaction types (lepton or hadron collisions, beam-dump experiments, neutrino-nucleus scattering, heavy ion collision, etc.).
In addition, an editor eases the implementation of personalizable detector models by the user.
A current disadvantage of \texttt{DDC} is the absence of implemented final-state-particle efficiencies, accounting for the detector sensitivity to various LLP decay products.
Nevertheless the code should allow enough flexibility for a later inclusion of this feature.

In the following section, we describe the essential features of the far-detector simulator \texttt{DDC}\footnote{We emphasize that our tool can in principle be used for the computation of signal event rates of displaced decays in the case of near detectors as well. In this case, the implementation of cuts (as allowed by \texttt{DDC}) becomes imperative in order to suppress the background.}.
We then apply this framework to several LLP scenarios in section~\ref{sec:benchmarks}, in order to illustrate the functionality of this tool (and validate our implementation of detector models).
A brief summary is provided in section~\ref{sec:conclusions}.

\section{The tool}\label{sec:tool}

\subsection{General features}
The C++ program \texttt{DDC} provides a framework based on \texttt{HepMC} and \texttt{Pythia} that is able to simulate detector response (up to a certain extent) for collider experiments looking for LLPs. Its basic input consists in:
\begin{itemize}
	\item a PDG code~\cite{MC_PID} representing a (long-lived) particle suspected to decay in a detector far from the interaction point, together with its mass, lifetime, and `visible' branching ratio; multiple LLPs are supported as input, as long as they are produced independently (and not in a chain);
	\item a set of Monte-Carlo events describing the production cross-section of this particle at a collider; \texttt{LHEF} and \texttt{HepMC} formats are accepted; alternatively, events can be generated internally via a call to \texttt{Pythia}: a \texttt{Pythia} input file in CMND format is then needed; finally, a total cross-section normalizing the set of events is also expected;
	\item a list of known detector models corresponding to the detectors that the user wants to simulate, together with the considered integrated luminosities (default values are associated to each detector but can be overwritten).
\end{itemize}
The event generation is entirely left to the care of the user, as such a feature strongly depends on the model. Dedicated tools (\texttt{MadGraph}, \texttt{Pythia}, \texttt{Herwig}~\cite{Bahr:2008pv,Bellm:2019zci}) are available on the market. Event generation through an internal call to \texttt{Pythia} is nevertheless possible via the CMND input.  Initial and final-state QCD radiation (showering), as well as hadronization, can also be run in-code via \texttt{Pythia}. Yet, such functionalities are not those targeted by our tool design and we will not comment further about them here.
We should emphasize that in order to obtain reliable predicted results, it is imperative to simulate sufficiently many events such that not too few simulated LLPs travel inside the considered detector's direction (up to the azimuthal angle in some cases).

Given the input described above, \texttt{DDC} computes the decay probability within the volume of the tested detector models of all tracked particles (i.e.~LLPs identified by their PDG code) appearing in the simulated events. The cross-section normalization, `visible' branching ratios, and integrated luminosities then allow to derive a number of expected events.
Detector efficiencies are not taken into account in the current version --- the corresponding information is generally not available for far-detector experiments --- even though an average efficiency can be included within the `visible' branching ratio input and within the weighing factors associated to (local) detector volumes (see the description in section 2.3).
A full MC simulation of the detector response to LLP decays within its volume (from which local detection efficiencies would be extracted) is a priori beyond the scope of DDC at present.
In addition, the program allows for the implementation of event cuts (e.g.\ restricting the energy of the LLPs) for each detector model.

\subsubsection{Getting started}
The source code for \texttt{DDC} can be downloaded from \newline\url{https://github.com/wzeren/Displaced-Decay-Counter}\ .
\newline Its operation requires installed versions of \texttt{Pythia8}~\cite{Sjostrand:2014zea} and \texttt{HepMC}~\cite{Dobbs:2001ck}\footnote{The length and energy units are set to mm and GeV for \texttt{HepMC}.} as prerequisites: we refer to the corresponding manuals for installation.
A C++17 compiler is also needed. After unpacking \texttt{DDC}, the paths in \texttt{DDC/Makefile} should be adjusted to the local setup. The \texttt{make} command then produces the central executable \texttt{main} in \texttt{DDC/bin}.

\subsubsection{Input}
The user's interface in \texttt{DDC} is ensured by input files in \texttt{json} format. Templates are provided in \texttt{DDC/bin}. As three types of input are needed (as described above), three input files are employed:
\begin{enumerate}
	\item The cross-section information is provided in the first file, as presented in the template file \texttt{DDC/bin/inputEvents.dat}. It consists in a choice of format (LHE, CMND or HEPMC) and a path to the simulated events (or the \texttt{Pythia} CMND run card); the number of events to be considered as well as a total cross-section in femtobarn serving as normalization should also be provided. The syntax reads:
	\newline \verb|{"input":{"input_file_format":"...",|\newline\verb|"input_file_path":"...","nMC":...,|\newline\verb|"sigma":...}}|.
	\item LLP characteristics (PID, lifetime, mass, visible branching ratio) are stored in a second file under the format (\verb|...| represents numerical entries): \newline\verb|{"LLP":{"LLPPID":...,"ctau":...,|\newline\verb|"mass":...,"visibleBR":...}}|. 
	\newline Templates for single or multiple LLPs are provided in \texttt{DDC/bin/inputLLPs.dat} for a single LLP and \texttt{DDC/bin/multipleLLPs.dat} for multiple LLPs.
	\texttt{ctau} is in meter and \texttt{mass} in GeV.
	\item The detector models (i.e.~implementations known by the program, as described below) considered in the simulation should be listed in \texttt{DDC/bin/detectors.dat} under the format: \verb|{"detector_name":[switch,int_luminosity]}|. 
	\newline\verb|detector_name| should correspond to a known detector model of the program (a list of default models is described in the following section). \texttt{switch}$=0$ or $1$ determines whether the detector is considered in the simulation. \verb|int_luminosity| corresponds to the assumed integrated luminosity. A negative number would be overwritten with default operating values.
\end{enumerate}

\subsubsection{Operation and output}
After compilation and once a valid input is available, the \texttt{DDC/bin/main} executable can be run in command line as (we use the names of template input files as example)
\newline \verb|./main inputEvents.dat inputLLPs.dat results.txt|
\newline Results are then saved in \texttt{DDC/bin/results.txt}, and some logs on the availability of the detectors and a summary of the input information are printed in the directory \texttt{DDC/bin/Logs/}, providing the simulated acceptance and observed number of events for each detector and each LLP (the latter might be distinguishable through their decay modes).

\subsection{Code structure}
The code content of \texttt{DDC} is organized around three main tasks:
\begin{enumerate}
	\item read and interpret the input files;
	\item encode the properties of detector models;
	\item analyze the MC events against the former.
\end{enumerate}

The LLP and event information is processed from the main routine via the class \texttt{inputInterface}. A class object is constructed from the paths to the input files. The input is then extracted and stored using the \texttt{json} interface: the characteristics of the LLPs are in particular stored in objects of the sub-class \texttt{CLLP}.

The characteristics of the detector models are stored in objects of the \texttt{Detector} class. The latter and its sub-classes are described in detail in the following subsection. The central property of these detector models rests with their class function computing a probability of decay within the detector based on geometric considerations, once an LLP trajectory (represented by its polar angle) and a boosted decay length have been identified.

The analysis of MC events is processed within the \texttt{analysis} class. Such an object is created in the main routine from the stored input, giving access to various class subroutines, which allow testing the events against the detector models. The event file is read (or produced with Pythia).
For each event, LLPs are searched for by their PDG code within the event chain.
The LLPs are checked against a list of detector models, which is generated from comparison of the requested input with the stored detector models. The outcome of this analysis is then copied in the output file after application of the relevant normalization factors.

Finally, we provide a Doxygen~\cite{doxygen} documentation at the following webpage:
\newline \url{https://wzeren.github.io/Displaced-Decay-Counter}\ .

\subsection{Implementation of the detector models}

\subsubsection{Modeling the decay probability}
The differential probability distribution of decay for an unstable particle with lifetime $\tau$ is described in its rest-frame by an exponential distribution as a function of the proper time $t^*$ of the particle: $\frac{d{\cal P}^{\text{dec}}_*(t^*)}{dt^*}=\tfrac{1}{\tau}\exp[-\tfrac{t^*}{\tau}]$. For a free particle flying in the laboratory frame with velocity $\vec{v}$ away from its production point at the origin $\text{O}$, we may translate this decay law in terms of the distance $\ell$ flown from the origin:
\begin{equation}
	\frac{d{\cal P}^{\text{dec}}_{\text{lab}}(\ell)}{d\ell}=\tfrac{1}{\gamma||\vec{v}||\tau}\exp[-\tfrac{\ell}{\gamma\lVert \vec{v}\rVert\tau}]
\end{equation}
where $\gamma\equiv(1-||\vec{v}||^2)^{-1/2}$ is the relativistic factor. Consequently, the probability that the particle decays in a detector that comes across its trajectory between distances $\ell_1$ and $\ell_2$ ($\ell_1<\ell_2$) simply reads:
\begin{equation}\label{eq:decprob}
	P^{\text{dec}}_{\text{lab}}(\ell_1,\ell_2)=\exp[-\tfrac{\ell_1}{\gamma\lVert \vec{v}\rVert\tau}]-\exp[-\tfrac{\ell_2}{\gamma\lVert \vec{v}\rVert\tau}]\,.
\end{equation}

Observing this simple fact, the estimation of the detector acceptance reduces to the simple geometrical exercise of determining the distances at which the LLP trajectory intersects the outer layers of the detector.
Additivity forms a remarkable property of the probability law of Eq.~(\ref{eq:decprob}): given a volume $V$ disjointly covered by sub-volumes $V_i$, $V=\sqcup_i V_i$, the probability that the particle decays in $V$ amounts to the sum of probabilities that it decays in the sub-volumes $V_i$. We will constantly exploit this property by covering the detector volume by a collection of sub-volumes. Finally, we observe that Eq.~(\ref{eq:decprob}) only involves the outer surfaces of the detector volume. Each surface intersected by the particle trajectory contributes an exponential term depending only on the distance of the intersection from the interaction point; the sign of this contribution is determined by the orientation of the surface with respect to that of the trajectory.

We stress three underlying hypotheses in this picture of the interplay between LLPs and the detector:
\begin{enumerate}
	\item the LLP is produced at the interaction point, so that displaced vertices (resulting from the decays of primary LLPs into secondary LLPs) are not taken into account;
	\item the LLP lifetime is not modified by the interactions of this particle with the matter constituting the detector;
	\item the LLP flies in a straight line from the interaction point, so that the impact of the electromagnetic fields, in particular surrounding the interaction point, on charged LLPs is neglected.
\end{enumerate}

\subsubsection{Defining detectors in cylindrical geometry}\label{subsubsec:def_detector}
In order to minimize the time cost of event simulation, we design the (virtual) detectors in a fashion exploiting the cylindrical symmetry (of LLP production) around the beam axis.
Detectors (which need not observe the cylindrical symmetry in their design) are thus constructed from `shapes' in the cylindrical half-plane --- the latter being defined by the beam axis $(\text{O}z)$ and the radial coordinate $(\text{O}h)$ ---, to which an angular aperture $\delta\varphi$ is associated. In this way, LLPs emitted at any azimuthal angle are used in the calculation of the acceptance.
Should control over the cylindrical angle prove necessary in a later stage of development of DDC,
it can be restored by associating angular cuts to the detector volumes.
For a detector covering an azimuthal aperture of $\Delta\varphi$, this type of implementation saves a factor $\approx2\pi/\Delta\varphi$ in terms of event generation, as compared to a strict three-dimensional modelization.

A set of basic objects is defined in the program in order to facilitate this implementation of detectors in cylindrical geometry:
\paragraph{Oriented cylindrical segments} (\texttt{class CylSeg}) define our most basic geometrical structure.
It is defined by two coordinate sets in the cylindrical plane (\texttt{std::array<double,2> AA=\{zA,hA\};}) representing its extremities, together with a sign (stored as an integer), representing its orientation (incoming or outgoing). A decay probability function is attached to this class, taking as input the polar angle $\theta$ of the particle trajectory and the effective decay length $\gamma\lVert \vec{v}\rVert\tau$, and returning an exponential term weighed by the orientation $\varepsilon_{\cal{O}}=\pm1$ of the segment if the trajectory intersects the segment at a distance $\ell$ from the interaction point (returning $0$ otherwise): $\varepsilon_{\cal{O}}\exp\big[-\tfrac{\ell}{\gamma\lVert \vec{v}\rVert\tau}\big]$. Given that the segment is `straight' in the cylindrical plane, it means that we renounce implementing curved surfaces in these directions, and that the latter need to be approximated by polygonal shapes.
\paragraph{Cylindrical detector layers} (\texttt{class CylDetLayer}) correspond to a list of oriented cylindrical segments (\texttt{std::vector<CylSeg> SegList}) with a weighing factor ${\cal W}$ (\texttt{double}). The corresponding decay probability function returns the sum of the decay probability terms associated with the individual segments, weighed by ${\cal W}$: ${\cal W}\cdot{\displaystyle\sum_{i}}\varepsilon_{\cal{O}}\exp\big[-\tfrac{\ell_i}{\gamma\lVert \vec{v}\rVert\tau}\big]$. The list of segments is meant to delimit a shape in the cylindrical plane, while the weight ${\cal W}$ primarily represents the azimuthal aperture $\delta\varphi$ associated with the detector layer ${\cal W}_{\varphi}=\delta\varphi/2\pi$. In a more sophisticated description of detectors, a sensitivity contribution of the detector layer to specific decay products of the decaying particle could be incorporated within this weight, along the line: ${\cal W}_{\cal S}={\displaystyle \sum_{\cal X}}\text{BR}[{\cal X}]\varepsilon^{\cal S}_{\cal X}$ where $\text{BR}[{\cal X}]$ represents the branching fraction of the LLP into the final state ${\cal X}$. However, in the current implementation, we only consider the detector acceptance, so that ${\cal W}_{\cal S}$ reduces to $\text{BR}[\text{visible}]$ and can be extracted as a global factor. Cuts (discussed below) offer another handle on the detector efficiency.

The detector layer can be directly constructed from a list of preliminarily defined segments and a weight: 
\newline\null\hfill\texttt{CylDetLayer::CylDetLayer(std::vector<CylSeg> Seglist, double wgh)}\,.\hfill\null 
\newline As such a definition could however accommodate comparatively exotic constructions that would not amount to a realistic detector layer, we also provide a constructor taking a list of coordinates and the weight as input:
\newline\null\hfill\texttt{CylDetLayer::CylDetLayer(std::vector<std::\newline array<double,2>> ptlist, double wgh)}\,.\hfill\null
\newline The constructor then builds the convex polygonal envelope of the considered list of points (\texttt{ptlist}), and  automatically determines the orientation of each derived segment.

The cylindrical detector layers are meant to be the building blocks of our detector modelization. They should be used to fill the volume of a detector. A simple (though rarely optimal) possibility would consist in defining detector layers of elementary dimensions $\delta z$, $\delta h$, $\delta\varphi$ and fill the detector volume with such elementary `bricks'. The function
\newline\texttt{CylDetLayer CylBrick(std::array<double,2> coord, double length,  double height, double apphi, double wgh)}
\newline can be used in such an approach: it defines a detector layer of rectangular shape in the cylindrical plane, of dimensions \texttt{length}$\times$\texttt{height}, centered on one set of coordinates \texttt{coord}. Other elementary shapes fulfilling this purpose are left to the user's imagination.

\paragraph{Detectors} (\texttt{class Detector}) combine an identifier (\texttt{std::string}), an integrated luminosity (\texttt{double}; this is a default value which can be overwritten in the input), and a list of cylindrical detector layers (\texttt{std::vector<CylDetLayer>}). The associated function \texttt{Detector::DetAcc(double th,double leff)} simply sums the weighed decay probabilities of individual layers for a particle emitted at a polar angle $\theta=$\texttt{th} and characterized by the decay length \texttt{leff}$=\gamma\lVert \vec{v}\rVert\tau$. A cut function is also associated with each detector. This object represents the higher level of description of the detector in cylindrical coordinates. Contrary to the case of smaller objects, we do not provide automatic building functions (from coordinates) for detectors. The user may refer to the built-in detectors that we provide for templates.

\subsubsection{Built-in detectors}
A list of virtual detectors is provided by default within our tool. These objects are meant to model existing designs of far detectors at the LHC\footnote{We stress that the tool is not at all restricted to applications in the LHC collisions.}. The mismatch in volumes between our implementation and the actual project (i.e.~the sum of all sub-volumes belonging to only the implementation, or only to the physical detector), normalized to the total volume of the detector offers an estimate of the accuracy of the description. In practice, compensation between fictitious and missing detector sub-volumes leads to a better precision at the level of the acceptance. Nevertheless, the provided detector models in cylindrical coordinates are often perfectible.

\begin{figure}[t]
	\begin{center}
		\includegraphics[width=0.5\textwidth]{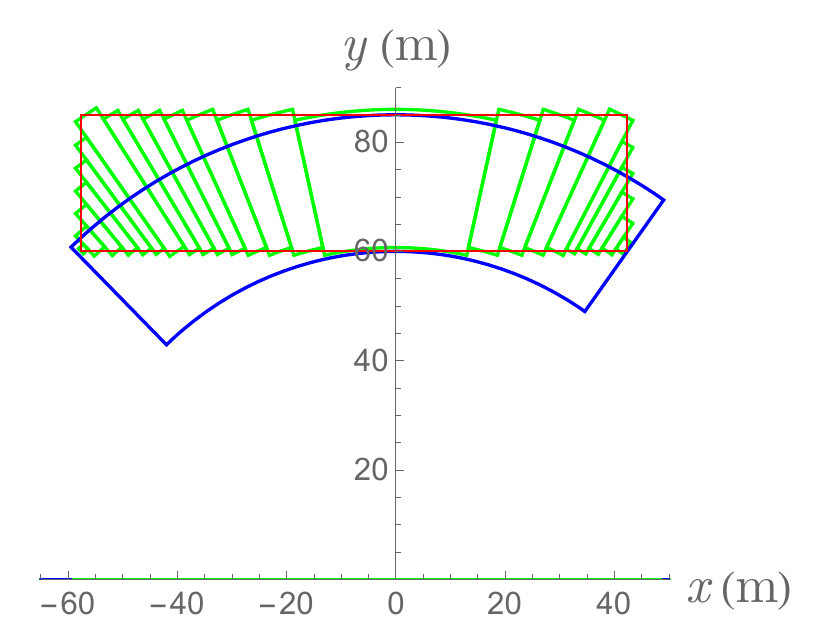} \includegraphics[width=0.5\textwidth]{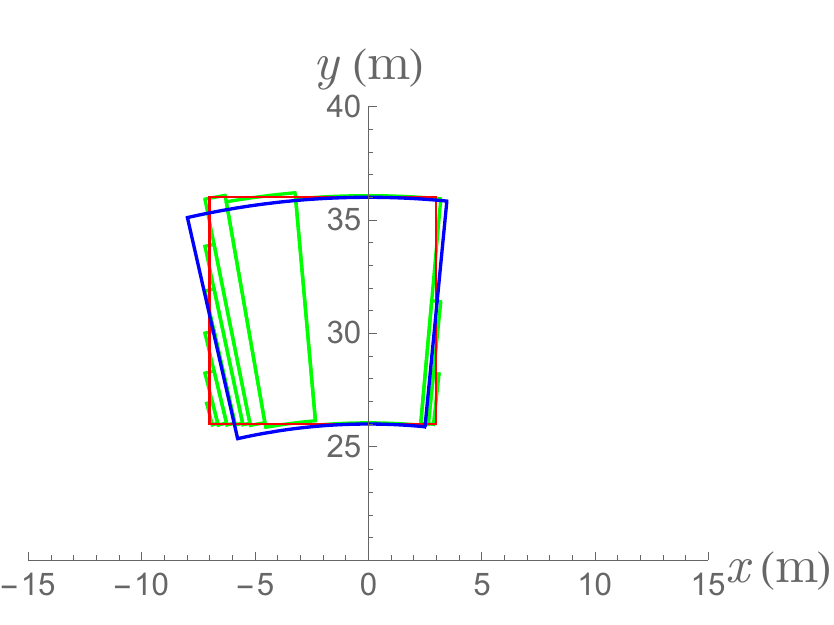}
		\caption{Comparison between the actual detector layouts (in red), the simple designs (in blue) and the more exact coverage (in green) that we provide in the polar plane.
			\newline{\em Upper}: MATHUSLA (red) vs.\ \texttt{MATHUSLA0} (blue) and \texttt{MATHUSLA1} (green);
			\newline{\em Lower}: CODEX-b (red) vs.\ \texttt{CODEXB0} (blue) and \texttt{CODEXB1} (green).
			\newline The axes $x$ and $y$ (in meters) span the polar plane (i.e.~are orthogonal to the collision axis $(Oz)$).
			\label{fig:detcomp}}
	\end{center}
\end{figure}

\paragraph{MATHUSLA} \cite{MATHUSLA:2020uve} is a box-shaped far-detector project of massive $100\,\text{m}\times100\,\text{m}\times25\,\text{m}$ dimensions for the CMS interaction point. Its depth is parallel with the beam axis, with near and far collinear sides at respectively $z_{\text{min}}=68$\,m and $z_{\text{max}}=168$\,m. The base and cover of the box stand at respectively $y_{\text{min}}=60$\,m and $y_{\text{max}}=85$\,m above the beam. The left- and right-hand side boundaries are slightly offset: $x_{\text{min}}=-42.41$\,m and $x_{\text{max}}=57.59$\,m. A nominal LHC integrated luminosity of $3000$\,fb$^{-1}$ is associated per default to this detector. We offer two models for this project:
\begin{itemize}
	\item \texttt{MATHUSLA0} is a basic but popular approximation consisting of a single detector layer delimited by the coordinates $z_{\text{min}}=68$\,m, $z_{\text{max}}=168$\,m, $h_{\text{min}}=60$\,m, $h_{\text{max}}=85$\,m, and the angular aperture $\Delta\varphi\approx1.39$\,rad. The mismatch in volume with the actual design is sizable, leading to an uncertainty of order $30\%$ at the level of the acceptance.
	\item \texttt{MATHUSLA1} is constructed out of $21$ cylindrical layers, aiming at a mismatch in volume below $4\%$ as compared to the actual design, and a precision on the acceptance at the percent level.
\end{itemize}
The projections of these models in the polar plane (orthogonal to the beam axis) are shown in the upper plot of Fig.~\ref{fig:detcomp}.

\paragraph{FASER and FASER2} \cite{FASER:2018eoc} are projected cylindrical detectors aligned with the beam axis in the far-forward region of the ATLAS interaction point ($480$\,m). The dimensions of FASER (FASER2) are given by a length of $1.5$\,m ($5$\,m) and a radius of $0.1$\,m ($1$\,m). The assumed integrated luminosities are $150$\,fb$^{-1}$ and $3000$\,fb$^{-1}$, respectively. Given that this geometry respects the cylindrical symmetry, there is no difficulty in modeling these detectors with a single detector layer. The corresponding virtual detectors that we provide are straightforwardly named \texttt{FASER} and \texttt{FASER2}.

\paragraph{CODEX-b} \cite{Gligorov:2017nwh} is a projected cubic detector of ($10$\,m)$^3$ dimensions at about $25$\,m from the LHCb interaction point and with two sides parallel to the beam axis. Its coordinates read: $z_{\text{min}}=5$\,m, $z_{\text{max}}=15$\,m, $y_{\text{min}}=26$\,m, $y_{\text{max}}=36$\,m, $x_{\text{min}}=-7$\,m and $x_{\text{max}}=3$\,m (we have exchanged the $x$- and $y$-axes). The integrated luminosity is assumed to reach $300$\,fb$^{-1}$. Again, both a simple and a more exact model are proposed:
\begin{itemize}
	\item \texttt{CODEXB0} is built out of a single detector layer extending between $5$ and $15$\,m in the $z$-direction, $26$ to $36$\,m in the $h$-direction and taking an angular aperture of $\Delta\varphi\approx0.32$\,rad. In view of the compact dimensions of CODEX-b, this approximation works rather well: the mismatch in volume with the actual layout is of order $10\%$, and the precision of the description should amount to a few percent.
	\item \texttt{CODEXB1} is constructed out of ten detector layers and aims at a geometrical accuracy better than $2\%$.
\end{itemize}
The projections of these models in the polar plane (orthogonal to the beam axis) are shown in the lower plot of Fig.~\ref{fig:detcomp}.

\paragraph{AL3X} \cite{Gligorov:2018vkc} is a proposed cylindrical detector with inner radius $0.85$\,m, outer radius $5$\,m and a length of $12$\,m, extending from $z_{\text{min}}=5.25$\, down the beam axis of the interaction point in the ALICE cavern. Again, the geometry does not raise difficulties and the model \texttt{AL3X} consists of a single detector layer. The default integrated luminosity is set to $250$\,fb$^{-1}$.

\paragraph{ANUBIS} \cite{Bauer:2019vqk} is a proposal for a detector of cylindrical shape, with diameter $18$\,m, length $56$\,m, placed in one of the service shafts about $24$\,m above the ATLAS beam line\footnote{A new geometrical design of ANUBIS has been since recently under discussion; see Refs.~\cite{ANUBIS_talk_slides,Satterthwaite:2839063}. We stick to the original proposal here and note that the new design can also be easily implemented in the code.}, which is parallel with the base of the cylinder.
The central axis is at $z_C=14$\,m. The targeted integrated luminosity is $3000$\,fb$^{-1}$. We propose two models:
\begin{itemize}
	\item \texttt{ANUBIS0} is a simple approximation with three superposed cylindrical layers of rectangular shape ($\Delta z=18$\,m, $\Delta h\approx18.67$\,m) and azimuthal apertures $\Delta\varphi_1\approx0.53$\,rad, $\Delta\varphi_2\approx0.34$\,rad and $\Delta\varphi_3\approx0.25$\,rad. The mismatch in volume is of order $50\%$.
	\item \texttt{ANUBIS1} is constructed with `cylindrical bricks' of extensions $\delta z=\delta h=1$\,m and angular precision $\delta\varphi\sim4\cdot10^{-3}$\,rad that fill the actual detector volume. The geometrical accuracy should improve onto $3\%$.
\end{itemize}

\paragraph{MAPP1 and MAPP2} \cite{Pinfold:2019nqj} are projected sub-detectors of the MoEDAL experiment at LHCb. Both have trapezoidal shapes with no side aligned with the beam axis. The planned integrated luminosities are respectively $30$\,fb$^{-1}$ and $300$\,fb$^{-1}$. As the cylindrical geometry is not optimal to describe such volumes, we simply employ the `brick' strategy, filling the volumes with cylindrical objects of elementary size, with a targeted accuracy of about $4\%$. The resulting models are simply called \texttt{MAPP1}  and \texttt{MAPP2}.

\paragraph{FACET} \cite{Cerci:2021nlb} is a cylindrical detector project placed at $101$\,m of the CMS interaction point. Its radius is $0.5$\,m and its length $18$\,m. This straightforward geometry is implemented as a single detector layer in the \texttt{FACET} model. It covers the polar angle between 1 and 4 mrad.
The expected integrated luminosity is 3000 fb$^{-1}$.

\subsubsection{Implementing new detectors}
The user can implement new detectors, either in order to investigate the potential of new geometries or simply to improve on the features of the built-in detectors. To this end, an assistant can be invoked. The corresponding executable needs first be created by running \verb|make DetEditor| in the main folder. Then, DetEditor is available in the \texttt{DDC/src} folder, and can be called from the command line as \texttt{./DetEditor}. Its function is to prepare a skeleton code and link it to the rest of the program. The only needed input for the assistant is the name of the new detector. The user can then access the barebone script in \texttt{DDC/src/Detectors} and edit the properties of the considered detector.
See sec.~\ref{subsubsec:def_detector} for details on a detector's geometrical definition in the code.

\subsubsection{Implementing cuts}
In order to better model the detector response, it is possible to implement cuts on events containing LLPs. A cut function is attached to each detector in the \texttt{DDC/src/Detectors} folder. It is by default trivial but can be edited by the user. The input of this function is a collider event in \texttt{HepMC} format, leaving a maximal flexibility on the possible types of conditions (possibly beyond mimicking the detector response).
Nevertheless, we note that, while the implementation of cuts is critical in the description of detectors located close to the main interaction points, it remains largely incidental in modelling the response of far detectors at the current stage of their development.~\footnote{The corresponding study in Ref.~\cite{Curtin:2023skh} in the case of the MATHUSLA experiment in particular demonstrates the very limited impact of detector cuts on the phenomenological results.}

\section{Physics benchmarks}\label{sec:benchmarks}

In this section, we illustrate the functionality of the tool \texttt{DDC} via three simple LLP models that we test against the detector projects described in the previous section.
As we consider only detector models planned for the LHC, events are produced in proton-proton collisions with 13 or 14~TeV center-of-mass energy.
Nevertheless, we stress that other types of colliders and detectors can be encoded for simulation within \texttt{DDC} in a straightforward manner.
We use \texttt{Pythia8.245} for the numerical studies in this section.
The default settings are used except that we turn off multi-parton interactions in order to speed up the simulation process without affecting the results.

\subsection{Light neutral fermion produced in bottom decays}
The first scenario concerns a light neutral exotic fermion (denoted as $N$ here) with mass under $5$\,GeV. Such a new-physics state can indeed escape collider constraints as long as it is essentially singlet under the electroweak gauge group (so that it is not produced in over-abundant proportions in $Z$-boson decays and is not accompanied by a comparatively light charged partner).
Astrophysical and cosmological limits do not apply as long as the considered particle is not stable (or very long-lived).
The decays of the light fermion often proceed through lepton- and/or baryon-number violating processes.
Examples of such a light decaying fermion include a massive sterile neutrino (see e.g.~Ref.~\cite{DeVries:2020jbs} for a study of long-lived sterile neutrinos produced in meson decays at the LHC) or a light bino in R-parity violating SUSY (RPV-SUSY) models (existing studies include for instance Refs.~\cite{Dercks:2018eua,Dercks:2018wum,Dreiner:2020qbi}).
The production of such a particle at colliders may proceed through numerous channels, lepton- and baryon-number conserving or violating.

Here, we focus on a lepton-number violating, baryon-number conserving production in bottom hadron decays (under the understanding that $N$ carries no charge under lepton number).
For simplicity, we focus on production modes mediated by partonic operators of the form $(\overline{N}\Gamma_l e)(\overline{b}\Gamma_q u)$ or $(\overline{N}\Gamma_l \nu_e)(\overline{b}\Gamma_q d)$, involving leptons ($e$, $\nu$) and quarks ($u$, $d$) of the first generation and where  $\Gamma_l$ and $\Gamma_q$ represent (model-dependent) Dirac structures.
Bottom hadrons $H_b$ may then decay into $N$ and a lepton, with possible additional light hadrons $H_q$.
The most relevant $H_b$'s for $N$-production are the comparatively long-lived $B$-mesons and $\Lambda_b^{0}$ baryons (other $H_b$'s have a much larger SM-like width, making an exotic disintegration likely uncompetitive) and conservation of the angular momentum forbids the two-body $\Lambda_b^{0}\to N\nu$.
Thus, under the further hypothesis of dominant two-body decays, $N$-production is largely captured by the disintegration of $B$-mesons $B^+\to N e^+$, $B^0\to N\nu$, likely comparable in magnitude as long as the new-physics controlling the short-distance effect is $SU(2)_L$-conserving (here, we specialize in the case where $u$ and $d$ are left-handed fields). 
We assume that the decay of the exotic fermion is not correlated with the production mode, so that the lifetime can be set independently from the production cross-section (this is typically the case in a R-parity-violating SUSY model, for instance, where different couplings mediate these different processes).
Possible $N$ decays include baryonic, semi-leptonic, and (radiative) leptonic channels~\cite{Domingo:2022emr}, and need not be specified here, as long as we ignore the sensitivity of the detectors to specific decay modes.
Finally, we specialize in the case of a mass $m_N=1$\,GeV for the exotic fermion, and consider various values of the proper lifetime $c\tau_N$, with an assumed visible branching ratio of $100$\,\%.

\begin{figure}[t]
	\begin{center}
		\includegraphics[width=0.5\textwidth]{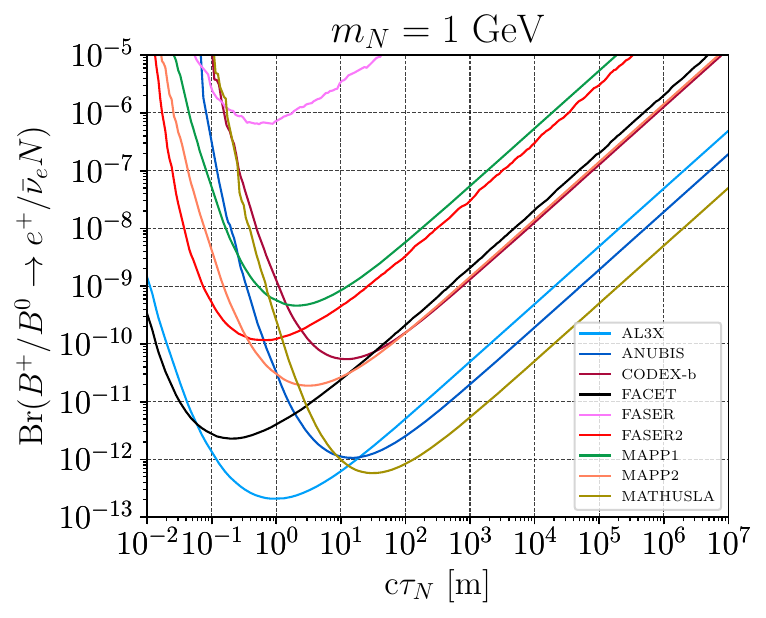} 
		\includegraphics[width=0.5\textwidth]{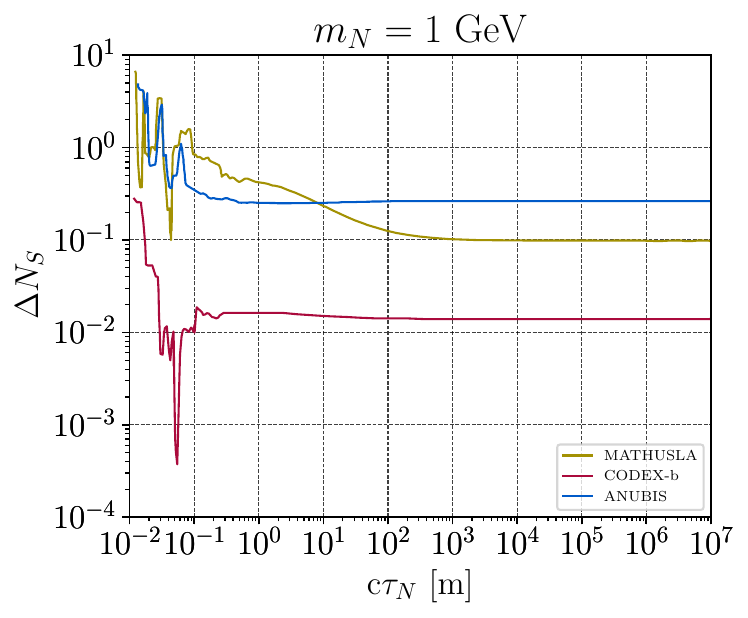}
		\caption{\textit{Upper}: Ideal exclusion bounds of far detectors at 95\% confidence level for a long-lived neutral fermion produced in $B^+$ and $B^0$ meson decays at the LHC far detectors.
			The $3$-signal-event boundaries are shown in the plane Br$(B^+/B^0 \to e^+/\bar{\nu}_e N)$ vs.~$c\tau_N$, when the neutral fermion mass  is fixed at 1 GeV. 
			(They represent upper limits on the branching ratios.)
			\textit{Lower:} Comparison between the implementations of simple approximations and more sophisticated implementations, in terms of the defined variable $\Delta N_S$, for \texttt{ANUBIS0} and \texttt{ANUBIS1}, \texttt{CODEXB0} and \texttt{CODEXB1}, and \texttt{MATHUSLA0} and \texttt{MATHUSLA1}.
		}
		\label{fig:B-meson}
	\end{center}
\end{figure}

The generation of the parton-level $pp\to b\bar{b}$ events and the ensuing hadronization processes at the center-of-mass energy $\sqrt{s}=14$ TeV are performed with \texttt{Pythia8} (here used internally within \texttt{DDC}).
The $B^\pm$ meson production cross sections over the whole $4\pi$ solid angle at the 14-TeV LHC is estimated to be about $4.87\times 10^{11}$ fb~\cite{DeVries:2020jbs}. 
The branching ratio of the exotic $B$-meson decay (into $e^\pm N$ or $\nu_e N$) is set to the fictitious value of $1$ in the event generation (for an efficient use of computer resources) and the production cross-section must be a posteriori re-scaled to account for the actual branching ratio.
The events are then analysed with \texttt{DDC} for multiple values of $c\tau_{N}$\footnote{For simulating ten thousand events for a benchmark parameter with one thread of a 10$^{\text{th}}$-generation i5 Intel CPU on a laptop, it takes about 12 seconds. The most computational resources are used by \texttt{Pythia8}.}.
We thus obtain the predicted numbers of signal events for each of the considered detectors at chosen integrated luminosities.
Here as well as in the following subsections, we model the MATHUSLA, CODEX-b, and ANUBIS experiments with the implementations \texttt{MATHUSLA1}, \texttt{CODEXB1}, and \texttt{ANUBIS1} described above.
The results are presented in the upper plot of Fig.~\ref{fig:B-meson}: the boundaries of the region with more than 3 signal events (95\% confidence level in the absence of background) are plotted in the plane spanned by Br$(B^+/B^0 \to e^+/\bar{\nu}_e  N)$ and $c\tau_{N}$ for a fixed $N$ mass of 1 GeV, considering the listed integrated luminosities.
The respective relevance of each experiment in terms of parameter-space coverage in this benchmark can be appreciated directly by reading the plot.
In the lower plot of Fig.~\ref{fig:B-meson} we analyze in which proportions the predictions depend on the modelization of the detectors.
We define the variable $\Delta N_S=|N_S^1 - N_S^0|/N_S^0$, where $N_S^{(0/1)}$ denotes the corresponding signal-event numbers, in order to quantify the relative difference between the simplistic and more sophisticated modelizations, namely \texttt{ANUBIS0} and \texttt{ANUBIS1}, \texttt{CODEXB0} and \texttt{CODEXB1}, and \texttt{MATHUSLA0} and \texttt{MATHUSLA1}.
The wiggles in the small $c\tau_N$ limit are due to insufficient simulation statistics in this regime. Expectedly, the compact dimension of \texttt{CODEXB} ensure a good percent-level performance of \texttt{CODEXB0}, while the discrepancies are larger for the more massive \texttt{MATHUSLA} ($\sim10\%$) and \texttt{ANUBIS} ($\sim30\%$). Still, we observe that the order of magnitude is properly captured by the simple approximations, which legitimates their common use in the literature.

Models of long-lived heavy neutral leptons (HNLs) or light RPV neutralinos produced in $B$-decays have been previously studied in e.g.~Refs.~\cite{Helo:2018qej,Dercks:2018eua,Dercks:2018wum,Hirsch:2020klk,Dreiner:2020qbi} from the perspective of far-detector proposals. The results of \texttt{DDC} were compared to those obtained in these papers and showed good agreement, up to traceable discrepancies in the choice of $B$-meson parameters or the inclusion of three-body decays, for instance. In the case of CODEX-b, ANUBIS, and MATHUSLA, simple approximations were routinely employed, which we thus tested against the \texttt{CODEXB0}, \texttt{ANUBIS0}, and \texttt{MATHUSLA0} designs. The compatibility of our results with those of earlier works argues in favor of the validity of the implemented detector models within \texttt{DDC}.

Below, we focus on the specific case of HNLs in order to compare our results with those obtained with \texttt{FastSim}.

\subsubsection{Heavy neutral lepton in the minimal scenario}\label{subsubsec:hnl_minimal}

In the minimal scenario for Majorana neutrinos, the SM is extended with three HNLs which mix with the active neutrinos.
The interaction Lagrangian for the HNLs and the SM electroweak gauge bosons is:
\begin{eqnarray}\label{eq:hnl-lagrangian}
	{\cal L} &=& \frac{g}{\sqrt{2}}\,\sum_{\alpha, j} 
	U_{\alpha j}\ \bar \ell_\alpha \gamma^{\mu} P_L N_{j} W^-_{L \mu} 
	\nonumber \\
	&&+\frac{g}{2 \cos\theta_W}\, \sum_{\alpha, i, j}V^{L}_{\alpha i} V_{\alpha j}^*  
	\overline{N_{j}} \gamma^{\mu} P_L \nu_{i} Z_{\mu},
\end{eqnarray}
where $i,j=1,2,3$, and $\ell_\alpha$, with $\alpha=e, \mu, \tau$, denoting the charged leptons of the SM.
$U_{\alpha j}$ is the mixing-matrix element relating the (dominantly singlet) HNL $N_j$ to its subdominant $SU(2)_L$ component $\nu_{\alpha}$.
$V^{L}$ labels the PMNS matrix.

For phenomenological purposes, a ``3+1'' scenario is commonly assumed where the sterile neutrino mixes with only one generation of the active neutrinos.
Here, we restrict ourselves to the case of a single HNL $N$ that mixes with the electron neutrino only, and consider its production in the rare decays of both charm mesons ($D^0, D^\pm$, and $D_s^\pm$) and bottom mesons ($B^0, B^\pm$, and $B_s^0$).
We follow Ref.~\cite{Gunther:2023vmz} for the evaluation of the total number of such mesons produced in LHC collisions, as well as that of their decay rates into HNL's and the subsequent HNL decay rates into SM particles.
The mixing angle plays a central role in enabling the latter two.
All the two- and three-body semi-leptonic and leptonic decay channels of the HNL are considered (and computed at leading order in chiral perturbation theory) for the calculation of its total decay width, 
and among all the decay channels of the HNL, only the tri-neutrino ones are counted as the invisible final states.

\begin{figure}[t]
    \centering
    \includegraphics[width=0.495\textwidth]{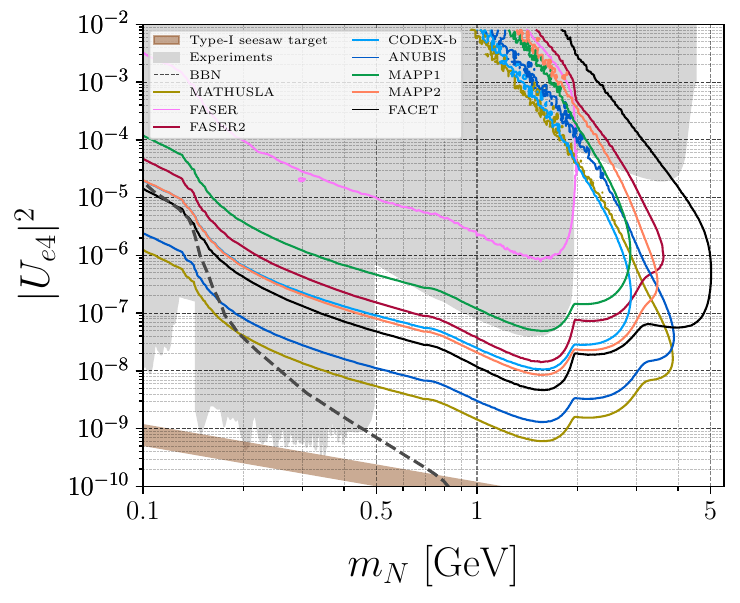}\\
    \includegraphics[width=0.495\textwidth]{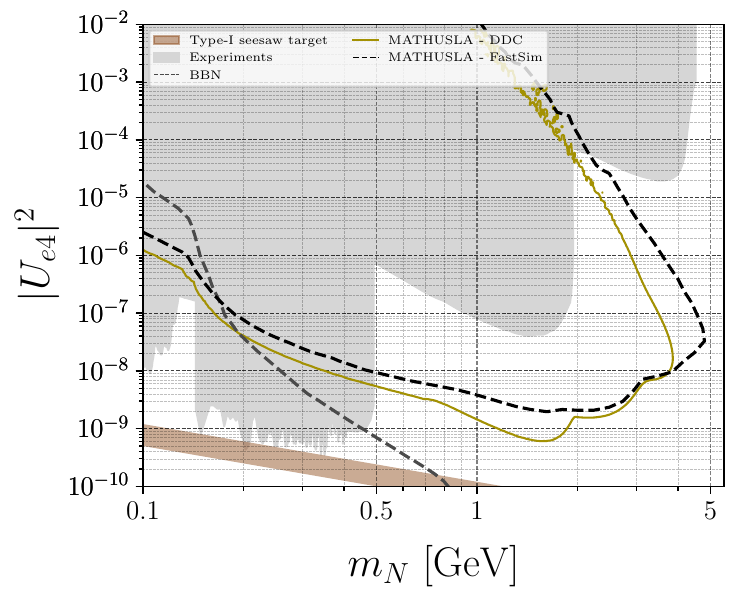}
    \caption{\textit{Upper}: Sensitivity reach of all the far detectors considered in this work at 95\% confidence level derived with \texttt{DDC}.
    The results are displayed in the parameter space of the HNL model spanned by the mass~$m_N$ and mixing-matrix element squared~$|U_{e4}|^2$.
    \textit{Lower}: Comparison of the sensitivity reach achieved by \texttt{DDC} in the case of MATHUSLA with corresponding results obtained in Ref.~\cite{Curtin:2023skh} with \texttt{FastSim}.
    The gray area in the background represents the regions currently excluded experimentally~\cite{Fernandez-Martinez:2023phj,PIENU:2017wbj,Bryman:2019bjg,NA62:2020mcv,T2K:2019jwa,Barouki:2022bkt,DELPHI:1996qcc}.
    The lower bounds from Big Bang Nucleosynthesis (BBN)~\cite{Sabti:2020yrt} is shown as a gray dashed curve, and the type-I seesaw target region for $m_{\nu_e}$ between 0.05 eV and 0.12 eV~\cite{Canetti:2010aw,Planck:2018vyg} is displayed in the brown band in the lower left part of the plots.
    }
    \label{fig:hnl_minimal}
\end{figure}

In Fig.~\ref{fig:hnl_minimal}, we present the expected sensitivity reach of far-detector experiments in the $(m_N, |U_{e4}|^2)$ plane, corresponding to the boundary of the region where 3 signal events can be detected within the exposure time of the experiment; these are identified with the exclusion bounds at 95\% confidence level in the expected absence of background events.
The upper plot contains the results obtained with \texttt{DDC} for MATHUSLA (with the \texttt{MATHUSLA1} implementation), FASER, FASER2, CODEX-b (with the \texttt{CODEXB1} implementation), ANUBIS (with the \texttt{ANUBIS1} implementation), MAPP1, and MAPP2.
Furthermore, in the background, we indicate the current upper bounds extracted from Refs.~\cite{Fernandez-Martinez:2023phj,PIENU:2017wbj,Bryman:2019bjg,NA62:2020mcv,T2K:2019jwa,Barouki:2022bkt,DELPHI:1996qcc} (the light gray area), the lower bounds computed from considerations relative to the Big Bang Synthesis (BBN)~\cite{Sabti:2020yrt} (the gray dashed curve), as well as the target region where a type-I seesaw is compatible with the active neutrino masses between 0.05 eV and 0.12 eV~\cite{Canetti:2010aw,Planck:2018vyg}.
In the lower plot, we compare the sensitivity reach of MATHUSLA computed with \texttt{DDC} with its counterpart employing \texttt{FastSim}~\cite{Curtin:2023skh}.
The latter is read from Fig.~6 of Ref.~\cite{Curtin:2023skh} and corresponds to the 4-signal-event curve.
We observe that both tools lead to largely compatible sensitivity limits.
The higher reach achieved by \texttt{FastSim} at large HNL mass can be essentially interpreted in terms of the more diverse production channels considered in Ref.~\cite{Curtin:2023skh}, in particular through the $B_c^\pm$ mesons (in addition to $W$ and $Z$-bosons as well as the $\tau$ leptons channels).
In the mass-range covered by both \texttt{DDC} and \texttt{FastSim}, the slightly weaker sensitivity predicted by \texttt{FastSim} can be attributed to a number of effects including e.g.~the precise number of mesons produced in the considered regime, as well as the finer treatment of the final-state particles and the LLP DV reconstruction within \texttt{FastSim} (see the discussion in Sec.~\ref{sec:intro}).
A detailed comparison goes beyond our purpose here, the general qualitative agreement being obvious.

\subsection{Long-lived complex scalar mixed with the SM Higgs boson and produced from \texorpdfstring{$B$}{}-meson decays}

In this example, we consider a complex neutral (electroweak singlet) scalar $S$, with mass $m_S$, that mixes with the SM Higgs according to an angle $\theta$.
Such a complex scalar consequently acquires a coupling to SM particles through $\theta$ and may then be produced in e.g.~rare decays of $B$-mesons such as $B\to K +S$, which we consider here.
Decays of the light scalar into SM final states are similarly expected.
For the calculation of the production and decay rates of the light scalar $S$, we follow the module provided in the FORESEE GitHub repository~\cite{FORESEE_github}.
For simplicity, we shall assume that the trilinear coupling between the complex scalar $S$ and two SM Higgs bosons $h$ is vanishing.
The following visible final states are considered in the decays of the light scalar: $S\to e^- e^+, \mu^- \mu^+, \pi\pi, KK, gg,$ and $ss$.

\begin{figure}[t]
    \centering
    \includegraphics[width=0.495\textwidth]{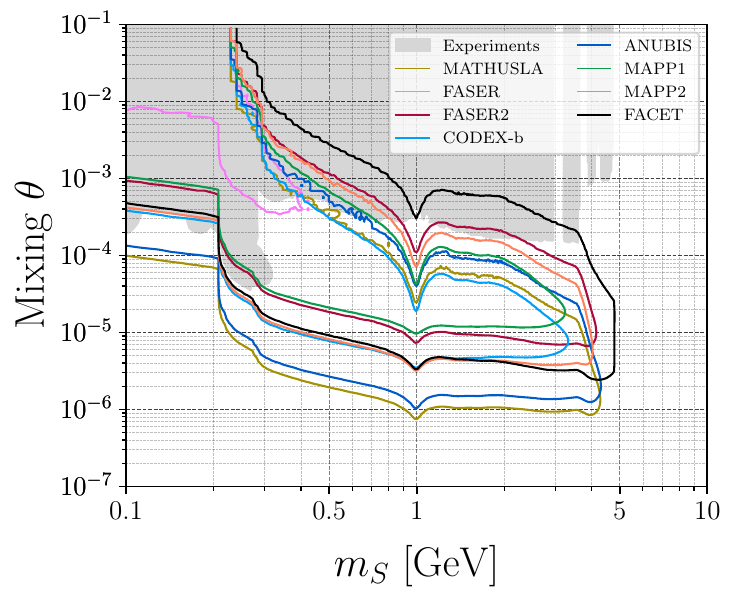}\\
    \includegraphics[width=0.495\textwidth]{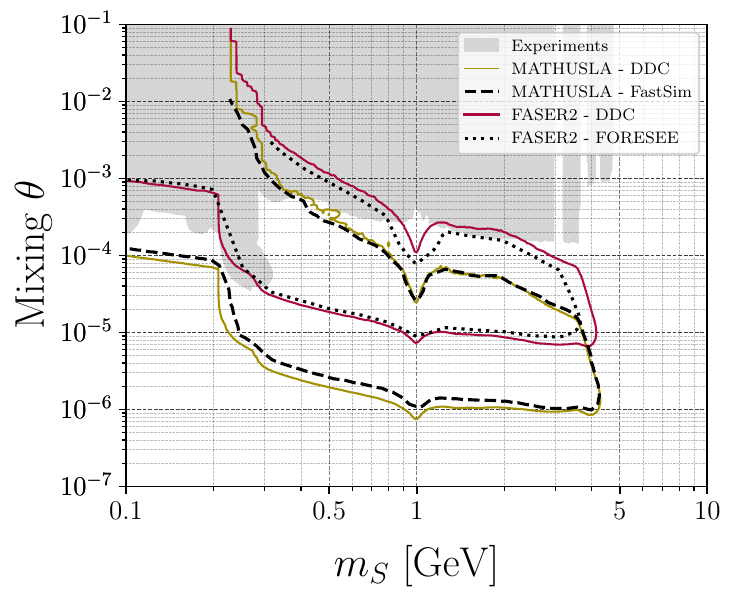}
    \caption{The projected sensitivity of the far-detected experiments are shown in the parameter space of the complex Higgs scalar model with mixing.
    The upper plot summarizes the results of \texttt{DDC}, while the lower plot performs a comparison with \texttt{FastSim} and \texttt{FORESEE} in the case of MATHUSLA and FASER2, respectively.
    The background gray area represents the current experimentally excluded region (see Ref.~\cite{Beacham:2019nyx} and the references therein).
    }
    \label{fig:light_scalar}
\end{figure}

Given the current bounds on $\theta$ applying in the relevant range of $m_S$, such light exotic scalars of the studied type are expected to be long-lived.
We perform the simulation and computation of the sensitivity prospects at LHC-based far-detector experiments using \texttt{DDC}.
The corresponding results are presented in Fig.~\ref{fig:light_scalar} in the $(m_S, \theta)$ plane.
Similarly to Fig.~\ref{fig:hnl_minimal}, we show the sensitivity reach of all the considered far detectors obtained with \texttt{DDC} in the upper plot, while the lower plot is dedicated to a comparison of our results with those obtained with \texttt{FORESEE} (3 signal events) and \texttt{FastSim} (4 signal events) in the case of FASER2 and MATHUSLA, respectively.
The MATHUSLA sensitivity curve at 4 signal events corresponding to \texttt{FastSim} is borrowed from Fig.~5 of Ref.~\cite{Curtin:2023skh}.
The gray area corresponds to the present bounds in the model-parameter space (see Ref.~\cite{Beacham:2019nyx} and the references therein).

Again, we observe a qualitatively good agreement among the sensitivity results derived with \texttt{DDC}, \texttt{FastSim}, and \texttt{FORESEE}, which demonstrates the efficiency of the approach based on detector acceptance that we use in our tool.
The minor differences on the exact position of the boundaries for the sensitivity regions may again be attributed to the detailed features implemented in each tool. In particular, \texttt{FORESEE}'s requirement on the energy of the LLP, $E_\text{LLP}>100$ GeV (to ensure a negligible background~\cite{Kling:2021fwx,FASER:2018ceo,FASER:2018bac}), results in a slightly more conservative coverage of the parameter space.
As explained before, similar features can be implemented within \texttt{DDC} using e.g.~its cut function, but we regard such a degree of refinement largely unnecessary at the current stage.

\subsection{Long-lived scalar in an extended Higgs sector}

\begin{figure}[t]
	\begin{center}
		\includegraphics[width=0.5\textwidth]{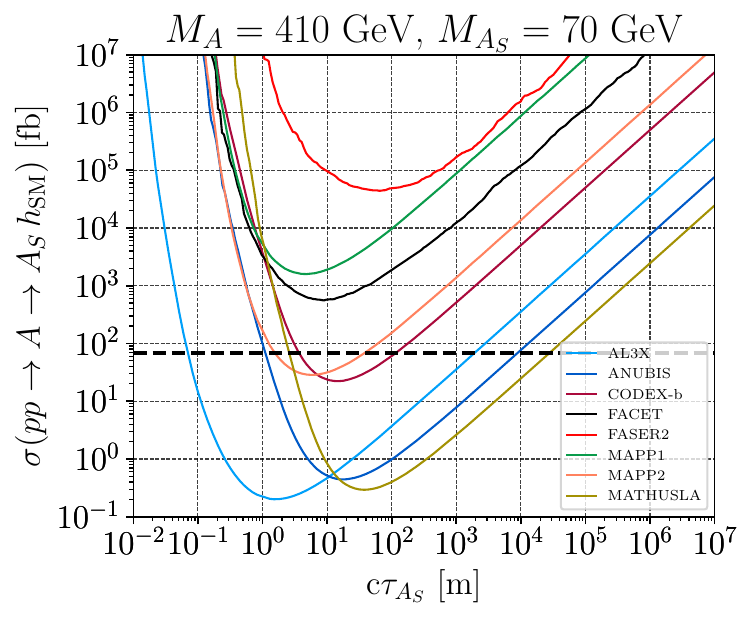} 
		\includegraphics[width=0.5\textwidth]{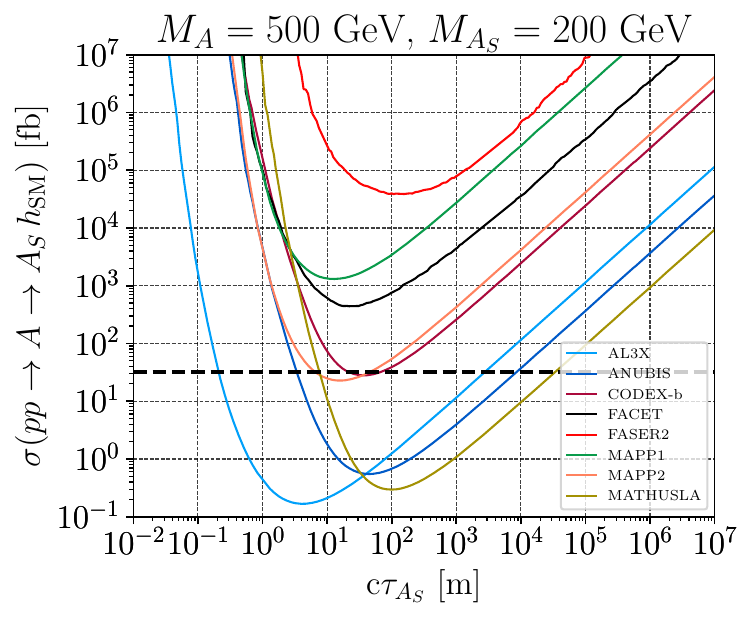} 
		\caption{Sensitivity plots for the extended Higgs models in the plane defined by the doublet production cross-section $pp \to A\to A_S h_{\text{SM}}$ and the singlet proper lifetime $c\tau_{A_S}$.
			The Higgs masses are chosen as $(M_A,M_{A_S})=(410,70)\,\text{GeV}$ in the upper plot and $(500,200)\,\text{GeV}$ in the lower plot.
			The corresponding cross-section values in Ref.~\cite{Ellwanger:2022jtd} are highlighted with a horizontal black dashed line.
			FASER has too weak a sensitivity to this scenario to appear in these plots.
		}
		\label{fig:scalar_higgs}
	\end{center}
\end{figure}

As a third example, we consider an extended Higgs sector, containing new doublet-dominated states $H$, $A$, $H^{\pm}$ at about $0.5$\,TeV and a lighter, mostly singlet CP-odd scalar $A_S$, with mass in the $100$\,GeV range.
In the limit where this pseudoscalar is purely singlet, its couplings to SM particles are suppressed, leading to a long lifetime, as well as a vanishing production rate in direct channels. However, some couplings of $A_S$ to BSM electroweakly-charged particles may remain sizable.
For instance, the trilinear couplings $A_S-A-h_{\text{SM}}$ and its $SU(2)$-conjugates $A_S-H-G^0$, $A_S-H^{\pm}-G^{\mp}$ are $SU(2)$-conserving~\cite{Domingo:2022kfm} and mediate the heavy-Higgs decays $A\to A_S h_{\text{SM}}$, $H\to A_S Z$, and $H^{\pm}\to A_S W^{\pm}$.
These channels may then dominate the total width of $A$, $H$ and $H^{\pm}$, which are easily produced in gluon fusion ($ggf$) at hadron colliders, or in association with top and bottom quarks (but may be difficult to observe if decaying in the indicated fashion).
These channels determine the production channels of the singlet.
We refer the reader to Ref.~\cite{Ellwanger:2022jtd} and references therein for a discussion of such a configuration in the context of the Next-to-Minimal SUSY SM (NMSSM), or more generally, in models with two Higgs doublets and a singlet.
The lifetime of $A_S$ can be almost arbitrarily large in such a setup.
In the NMSSM, the width may be dominated by the diphoton channel, mediated by chargino loops.
When this mechanism is absent, or in the non-SUSY case, the $A_S$ decays, typically into SM fermion pairs, proceed through the vanishingly small doublet component of this state, or exploit larger Higgs-to-Higgs couplings in loop-induced channels.

Here, we assume that $A_S$ is long-lived and focus on the production mode $gg\to A\to A_S h_{\text{SM}}$.
In practice, alternative production channels may be competitive,
in particular $gg\to H\to A_S Z$, or associated production with third-generation quarks, but we restrict ourselves to the single pseudoscalar channel as a simple handle on this scenario.
Realistic cross-section values can be inferred from Table~2 of Ref.~\cite{Ellwanger:2022jtd}, considering the $pp\to A\to (h_{\text{SM}}\to\tau^+\tau^-)(A_S\to\gamma\gamma)$ search channel, after re-scaling the corresponding numbers by the branching ratios of the final Higgs states, Br$(h_{\text{SM}}\to\tau^+\tau^-)$ and Br$(A_S\to\gamma\gamma)$.
Our scenario indeed differs from that of Ref.~\cite{Ellwanger:2022jtd} in that $A_S$ no longer promptly decays (and we do not pay attention to the decays of $h_{SM}$).
In practice, we will use the two benchmark points $(m_A, m_{A_S}) = (410, 70)$ GeV and $(500, 200)$ GeV (listed among other values in Table~2 of Ref.~\cite{Ellwanger:2022jtd}), vary the $pp\to A\to A_S h_{\text{SM}}$ cross section freely and assume a 100\% visibility of the decay products of $A_S$.

Again, we invoke \texttt{Pythia8} for event generation at $\sqrt{s}=13$ TeV (in accordance with Ref.~\cite{Ellwanger:2022jtd}) and let \texttt{DDC} compute the expected signal-event numbers at the various LHC far detectors.
The results are displayed in Fig.~\ref{fig:scalar_higgs}.
Again, the boundaries of the 3-signal events are shown in the plane spanned by $\sigma(pp\to A\to A_S h_{\text{SM}})$ and~$c\tau_{A_S}$, highlighting the discovery potential of the far detectors for long-lived scalars produced by a single scalar resonance.
The black horizontal dashed lines mark the cross-section values corresponding to Table 2 of Ref.~\cite{Ellwanger:2022jtd}, which may be deemed realistic for a Higgs-inspired model.

We stress that the results of this benchmark can be easily generalized to other models where a single heavy resonance is produced via gluon fusion and decays to an LLP.

\subsection{Long-lived fermion from a pair of heavy resonances}

In the final example, we consider a RPV-SUSY-inspired scenario where the production of SUSY resonances is still dominated by R-parity-conserving effects, while small R-parity-violating couplings mediate the decays of a long-lived LSP (lightest SUSY particle) bino $\tilde{\chi}^0_1$.
The latter may be arbitrarily long-lived, depending on the magnitude of the RPV couplings.
The binos can be directly produced in pairs via s-channel contributions involving their mixings with higgsinos, or via t-channel diagrams involving squarks.
However, if higgsinos and squarks are very heavy, which we assume for simplicity, the direct production cross-section is correspondingly suppressed.
Instead, we consider an indirect production mechanism where right-handed sleptons, $\tilde{e}_R$, exist in the $\sim 0.5$\,TeV range, are produced in pairs in $pp$-collisions (via their electroweak interactions) and decay into a lepton and a $\tilde{\chi}^0_1$ (via R-parity conserving gaugino couplings).
We further assume a $100\%$ visibility for the bino decays and the mass of this particle is set to $1$\,GeV. We do not pay particular attention to limits from prompt searches at the LHC here, as the scenario is simply meant for illustration, but mono- and multi-lepton 
searches can be relevant in this context, as the long-lived neutralino would leave a clear missing energy signature in the ATLAS or CMS detectors.

\begin{figure}[t]
	\begin{center}
		\includegraphics[width=0.5\textwidth]{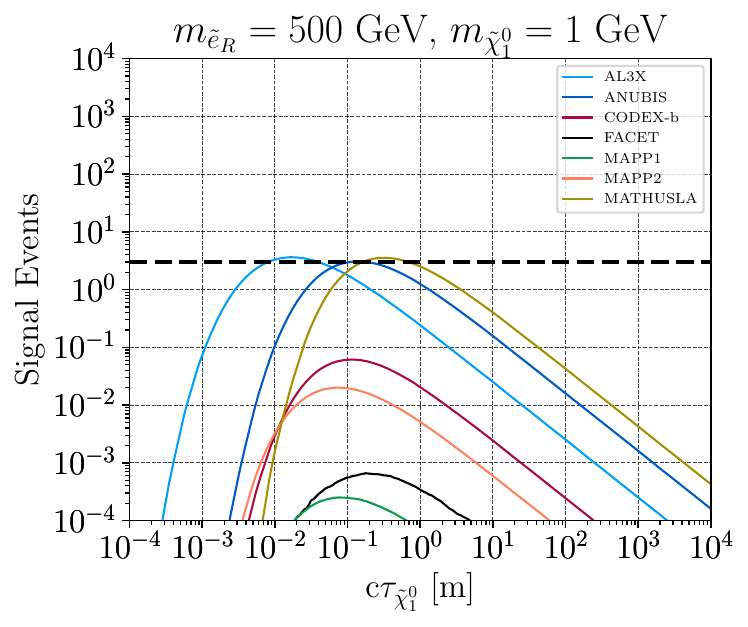} 
		\caption{Number of events for a light bino $m_{\tilde{\chi}^0_1}=1$  GeV, produced in the decays of selectrons pair produced, with $m_{\tilde{e}_L}=500$ GeV.
			The number of events are shown as a function of the lifetime of the bino.
			The black horizontal dashed line indicates the 3 signal event threshold.
			FASER and FASER2 have too weak sensitivity to this scenario to be displayed on this plot.
		}
		\label{fig:light_neutralino}
	\end{center}
\end{figure}

Once again, we generate events with \texttt{Pythia8} for the slepton pair production at $\sqrt{s}=13$ TeV in $pp$ collisions, with subsequent decays into binos and leptons. We normalize the cross-section to the next-to-leading and next-to-leading-logarithmic  augmented order, as provided in Ref.~\cite{pair_slepton_XS}. The sensitivity reach of the far-detectors is computed with \texttt{DDC} and presented in Fig.~\ref{fig:light_neutralino}. It shows the limited detection capabilities of far detectors in such a scenario, as the 3-signal event threshold is barely reached in the $c\tau_{\tilde{\chi}_1^0}\sim10^{-2}-1$\,m range.

This benchmark exemplifies the case of LLP production from a pair of heavy resonances, and we stress that it can be easily generalized to scenarios beyond the RPV-SUSY model that we considered here.

These simple scenarios have provided us with the opportunity of demonstrating straightforward functionalities 
of \texttt{DDC}.
The input files for all these sample benchmark studies, as well as the plotting scripts, the final sensitivity plots, and the data points, can be found in the folder \texttt{DDC/examples}.

\section{Conclusions}\label{sec:conclusions}

In this paper, we have described \texttt{Displaced Decay Counter (DDC)}, a versatile C++ program for the calculation of detector acceptances, relying on \texttt{Pythia8} and \texttt{HepMC}.
\texttt{DDC} is primarily meant as a simulation tool for long-lived-particle models and the currently booming far-detector proposals.
It accepts MC events as input in \texttt{LHEF} and \texttt{HepMC} format (a \texttt{Pythia8} run card can also be used for internal event generation).
The current version of \texttt{DDC} comes with a collection of far-detector designs, approximating several detectors currently planned or in construction at the LHC, and provides the user with an editor for the implementation of further far-detector models.

We have presented several benchmark scenarios to illustrate the usage of our tool.
The first one, the already well-studied case of light long-lived neutral particles produced in $B$-meson decays, gave us the opportunity to confront the performance of \texttt{DDC} to older results and validate the implemented detector models.
We then focused on the specific case of HNL's produced in the decays of charm and bottom mesons and noted the general agreement of our predictions with those derived for MATHUSLA with the tool \texttt{FastSim}.
Another scenario, with a long-lived scalar singlet mixing with the SM Higgs, was considered, mainly for comparison purposes with  \texttt{FORESEE} (in the case of the FASER2 experiment) and \texttt{FastSim} (in the case of MATHUSLA).
Again, we stressed the close agreement of the sensitivity prospects derived with \texttt{DDC}, with those obtained with these other tools, despite the currently less refined treatment of final states in our implemented detector models (strictly focusing on geometrical acceptance).
The further two models exemplify the cases of LLP production via a single or a pair of resonances in $pp$ collisions.
On a concluding note, we stress that \texttt{DDC} is highly versatile, offering modelling capabilities beyond those demonstrated here, and thus comes with a high potential for studying LLP models at colliders.

\bigskip
\bigskip

\section*{Acknowledgements}
We thank Charles Young for useful discussion.
No funds, grants, or other support was received.




\bibliography{refs}

\begin{thebibliography}{10}
\providecommand{\url}[1]{{#1}}
\providecommand{\urlprefix}{URL }
\providecommand{\doi}[1]{\url{https://doi.org/#1}}
\bibcommenthead

\bibitem{Curtin:2018mvb}
D.~Curtin, et~al., {Long-Lived Particles at the Energy Frontier: The MATHUSLA Physics Case}.
\newblock Rept. Prog. Phys. \textbf{82}(11), 116201 (2019).
\newblock \doi{10.1088/1361-6633/ab28d6}.
\newblock {\href{https://arxiv.org/abs/1806.07396}{{arXiv:1806.07396}}} {[hep-ph]}

\bibitem{Lee:2018pag}
L.~Lee, C.~Ohm, A.~Soffer, T.T. Yu, {Collider Searches for Long-Lived Particles Beyond the Standard Model}.
\newblock Prog. Part. Nucl. Phys. \textbf{106}, 210--255 (2019).
\newblock \doi{10.1016/j.ppnp.2019.02.006}.
\newblock [Erratum: Prog.Part.Nucl.Phys. 122, 103912 (2022)].
\newblock {\href{https://arxiv.org/abs/1810.12602}{{arXiv:1810.12602}}} {[hep-ph]}

\bibitem{Beacham:2019nyx}
J.~Beacham, et~al., {Physics Beyond Colliders at CERN: Beyond the Standard Model Working Group Report}.
\newblock J. Phys. G \textbf{47}(1), 010501 (2020).
\newblock \doi{10.1088/1361-6471/ab4cd2}.
\newblock {\href{https://arxiv.org/abs/1901.09966}{{arXiv:1901.09966}}} {[hep-ex]}

\bibitem{Alimena:2019zri}
J.~Alimena, et~al., {Searching for long-lived particles beyond the Standard Model at the Large Hadron Collider}.
\newblock J. Phys. G \textbf{47}(9), 090501 (2020).
\newblock \doi{10.1088/1361-6471/ab4574}.
\newblock {\href{https://arxiv.org/abs/1903.04497}{{arXiv:1903.04497}}} {[hep-ex]}

\bibitem{Agrawal:2021dbo}
P.~Agrawal, et~al., {Feebly-interacting particles: FIPs 2020 workshop report}.
\newblock Eur. Phys. J. C \textbf{81}(11), 1015 (2021).
\newblock \doi{10.1140/epjc/s10052-021-09703-7}.
\newblock {\href{https://arxiv.org/abs/2102.12143}{{arXiv:2102.12143}}} {[hep-ph]}

\bibitem{Knapen:2022afb}
S.~Knapen, S.~Lowette, {A guide to hunting long-lived particles at the LHC}  (2022).
\newblock {\href{https://arxiv.org/abs/2212.03883}{{arXiv:2212.03883}}} {[hep-ph]}

\bibitem{ATLAS:2022atq}
{Search for heavy neutral leptons in decays of $W$ bosons using a dilepton displaced vertex in $\sqrt{s}=13$ TeV $pp$ collisions with the ATLAS detector}  (2022).
\newblock {\href{https://arxiv.org/abs/2204.11988}{{arXiv:2204.11988}}} {[hep-ex]}

\bibitem{ATLAS:2022zhj}
G.~Aad, et~al., {Search for neutral long-lived particles in $pp$ collisions at $\sqrt{s}=13$ TeV that decay into displaced hadronic jets in the ATLAS calorimeter}  (2022).
\newblock {\href{https://arxiv.org/abs/2203.01009}{{arXiv:2203.01009}}} {[hep-ex]}

\bibitem{ATLAS:2022bll}
{Search for light long-lived neutral particles that decay to collimated pairs of leptons or light hadrons in $pp$ collisions at $\sqrt{s}=13$ TeV with the ATLAS detector}  (2022)

\bibitem{CMS:2022qej}
{Search for long-lived particles decaying to a pair of muons in proton-proton collisions at $\sqrt{s}$ = 13 TeV}  (2022).
\newblock {\href{https://arxiv.org/abs/2205.08582}{{arXiv:2205.08582}}} {[hep-ex]}

\bibitem{CMS:2021lcp}
{Search for long-lived particles decaying to displaced leptons in proton-proton collisions at $\sqrt{s}=13~\mathrm{TeV}$}  (2021)

\bibitem{ATLAS:2017tny}
M.~Aaboud, et~al., {Search for long-lived, massive particles in events with displaced vertices and missing transverse momentum in $\sqrt{s}$ = 13 TeV $pp$ collisions with the ATLAS detector}.
\newblock Phys. Rev. D \textbf{97}(5), 052012 (2018).
\newblock \doi{10.1103/PhysRevD.97.052012}.
\newblock {\href{https://arxiv.org/abs/1710.04901}{{arXiv:1710.04901}}} {[hep-ex]}

\bibitem{ATLAS:2020xyo}
G.~Aad, et~al., {Search for long-lived, massive particles in events with a displaced vertex and a muon with large impact parameter in $pp$ collisions at $\sqrt{s} = 13$ TeV with the ATLAS detector}.
\newblock Phys. Rev. D \textbf{102}(3), 032006 (2020).
\newblock \doi{10.1103/PhysRevD.102.032006}.
\newblock {\href{https://arxiv.org/abs/2003.11956}{{arXiv:2003.11956}}} {[hep-ex]}

\bibitem{ATLAS:2023oti}
{Search for long-lived, massive particles in events with displaced vertices and multiple jets in $pp$ collisions at $\sqrt{s} = 13$ TeV with the ATLAS detector}  (2023).
\newblock {\href{https://arxiv.org/abs/2301.13866}{{arXiv:2301.13866}}} {[hep-ex]}

\bibitem{ATLAS:2019gqq}
M.~Aaboud, et~al., {Search for heavy charged long-lived particles in the ATLAS detector in 36.1 fb$^{-1}$ of proton-proton collision data at $\sqrt{s} = 13$ TeV}.
\newblock Phys. Rev. D \textbf{99}(9), 092007 (2019).
\newblock \doi{10.1103/PhysRevD.99.092007}.
\newblock {\href{https://arxiv.org/abs/1902.01636}{{arXiv:1902.01636}}} {[hep-ex]}

\bibitem{Desai:2021jsa}
N.~Desai, F.~Domingo, J.S. Kim, R.R.d.A. Bazan, K.~Rolbiecki, M.~Sonawane, Z.S. Wang, {Constraining electroweak and strongly charged long-lived particles with CheckMATE}.
\newblock Eur. Phys. J. C \textbf{81}(11), 968 (2021).
\newblock \doi{10.1140/epjc/s10052-021-09727-z}.
\newblock {\href{https://arxiv.org/abs/2104.04542}{{arXiv:2104.04542}}} {[hep-ph]}

\bibitem{Araz:2021akd}
J.Y. Araz, B.~Fuks, M.D. Goodsell, M.~Utsch, {Recasting LHC searches for long-lived particles with MadAnalysis~5}.
\newblock Eur. Phys. J. C \textbf{82}(7), 597 (2022).
\newblock \doi{10.1140/epjc/s10052-022-10511-w}.
\newblock {\href{https://arxiv.org/abs/2112.05163}{{arXiv:2112.05163}}} {[hep-ph]}

\bibitem{Alguero:2021dig}
G.~Alguero, J.~Heisig, C.K. Khosa, S.~Kraml, S.~Kulkarni, A.~Lessa, H.~Reyes-Gonz\'alez, W.~Waltenberger, A.~Wongel, {Constraining new physics with SModelS version 2}.
\newblock JHEP \textbf{08}, 068 (2022).
\newblock \doi{10.1007/JHEP08(2022)068}.
\newblock {\href{https://arxiv.org/abs/2112.00769}{{arXiv:2112.00769}}} {[hep-ph]}

\bibitem{LLPrecastingRepo}
LLP-Recasting-Repository, {LLP Recasting Repository: \url{https://github.com/llprecasting/recastingCodes}} (2021)

\bibitem{delphes_pr}
C.~Wang, {Dedicated Delphes Module: \url{https://github.com/delphes/delphes/pull/103}} (2022)

\bibitem{FASER:2018eoc}
A.~Ariga, et~al., {FASER\textquoteright{}s physics reach for long-lived particles}.
\newblock Phys. Rev. D \textbf{99}(9), 095011 (2019).
\newblock \doi{10.1103/PhysRevD.99.095011}.
\newblock {\href{https://arxiv.org/abs/1811.12522}{{arXiv:1811.12522}}} {[hep-ph]}

\bibitem{Helo:2018qej}
J.C. Helo, M.~Hirsch, Z.S. Wang, {Heavy neutral fermions at the high-luminosity LHC}.
\newblock JHEP \textbf{07}, 056 (2018).
\newblock \doi{10.1007/JHEP07(2018)056}.
\newblock {\href{https://arxiv.org/abs/1803.02212}{{arXiv:1803.02212}}} {[hep-ph]}

\bibitem{Dercks:2018eua}
D.~Dercks, J.~De~Vries, H.K. Dreiner, Z.S. Wang, {R-parity Violation and Light Neutralinos at CODEX-b, FASER, and MATHUSLA}.
\newblock Phys. Rev. D \textbf{99}(5), 055039 (2019).
\newblock \doi{10.1103/PhysRevD.99.055039}.
\newblock {\href{https://arxiv.org/abs/1810.03617}{{arXiv:1810.03617}}} {[hep-ph]}

\bibitem{Dercks:2018wum}
D.~Dercks, H.K. Dreiner, M.~Hirsch, Z.S. Wang, {Long-Lived Fermions at AL3X}.
\newblock Phys. Rev. D \textbf{99}(5), 055020 (2019).
\newblock \doi{10.1103/PhysRevD.99.055020}.
\newblock {\href{https://arxiv.org/abs/1811.01995}{{arXiv:1811.01995}}} {[hep-ph]}

\bibitem{Hirsch:2020klk}
M.~Hirsch, Z.S. Wang, {Heavy neutral leptons at ANUBIS}.
\newblock Phys. Rev. D \textbf{101}(5), 055034 (2020).
\newblock \doi{10.1103/PhysRevD.101.055034}.
\newblock {\href{https://arxiv.org/abs/2001.04750}{{arXiv:2001.04750}}} {[hep-ph]}

\bibitem{Gorordo:2022rro}
T.~Gorordo, S.~Knapen, B.~Nachman, D.J. Robinson, A.~Suresh, {Geometry Optimization for Long-lived Particle Detectors}  (2022).
\newblock {\href{https://arxiv.org/abs/2211.08450}{{arXiv:2211.08450}}} {[hep-ph]}

\bibitem{Plows:2022gxc}
K.J. Plows, X.~Lu, {Modeling heavy neutral leptons in accelerator beamlines}.
\newblock Phys. Rev. D \textbf{107}(5), 055003 (2023).
\newblock \doi{10.1103/PhysRevD.107.055003}.
\newblock {\href{https://arxiv.org/abs/2211.10210}{{arXiv:2211.10210}}} {[hep-ph]}

\bibitem{Beltran:2023nli}
R.~Beltr\'an, G.~Cottin, M.~Hirsch, A.~Titov, Z.S. Wang, {Reinterpretation of searches for long-lived particles from meson decays}.
\newblock JHEP \textbf{05}, 031 (2023).
\newblock \doi{10.1007/JHEP05(2023)031}.
\newblock {\href{https://arxiv.org/abs/2302.03216}{{arXiv:2302.03216}}} {[hep-ph]}

\bibitem{Fernandez-Martinez:2023phj}
E.~Fern\'andez-Mart\'\i{}nez, M.~Gonz\'alez-L\'opez, J.~Hern\'andez-Garc\'\i{}a, M.~Hostert, J.~L\'opez-Pav\'on, {Effective portals to heavy neutral leptons}.
\newblock JHEP \textbf{09}, 001 (2023).
\newblock \doi{10.1007/JHEP09(2023)001}.
\newblock {\href{https://arxiv.org/abs/2304.06772}{{arXiv:2304.06772}}} {[hep-ph]}

\bibitem{Dreiner:2023gir}
H.K. Dreiner, D.~K\"ohler, S.~Nangia, M.~Sch\"urmann, Z.S. Wang, {Recasting Bounds on Long-lived Heavy Neutral Leptons in Terms of a Light Supersymmetric R-parity Violating Neutralino}  (2023).
\newblock {\href{https://arxiv.org/abs/2306.14700}{{arXiv:2306.14700}}} {[hep-ph]}

\bibitem{Buonocore:2018xjk}
L.~Buonocore, C.~Frugiuele, F.~Maltoni, O.~Mattelaer, F.~Tramontano, {Event generation for beam dump experiments}.
\newblock JHEP \textbf{05}, 028 (2019).
\newblock \doi{10.1007/JHEP05(2019)028}.
\newblock {\href{https://arxiv.org/abs/1812.06771}{{arXiv:1812.06771}}} {[hep-ph]}

\bibitem{Alwall:2011uj}
J.~Alwall, M.~Herquet, F.~Maltoni, O.~Mattelaer, T.~Stelzer, {MadGraph 5 : Going Beyond}.
\newblock JHEP \textbf{06}, 128 (2011).
\newblock \doi{10.1007/JHEP06(2011)128}.
\newblock {\href{https://arxiv.org/abs/1106.0522}{{arXiv:1106.0522}}} {[hep-ph]}

\bibitem{Alwall:2014hca}
J.~Alwall, R.~Frederix, S.~Frixione, V.~Hirschi, F.~Maltoni, O.~Mattelaer, H.S. Shao, T.~Stelzer, P.~Torrielli, M.~Zaro, {The automated computation of tree-level and next-to-leading order differential cross sections, and their matching to parton shower simulations}.
\newblock JHEP \textbf{07}, 079 (2014).
\newblock \doi{10.1007/JHEP07(2014)079}.
\newblock {\href{https://arxiv.org/abs/1405.0301}{{arXiv:1405.0301}}} {[hep-ph]}

\bibitem{Kling:2021fwx}
F.~Kling, S.~Trojanowski, {Forward experiment sensitivity estimator for the LHC and future hadron colliders}.
\newblock Phys. Rev. D \textbf{104}(3), 035012 (2021).
\newblock \doi{10.1103/PhysRevD.104.035012}.
\newblock {\href{https://arxiv.org/abs/2105.07077}{{arXiv:2105.07077}}} {[hep-ph]}

\bibitem{Jerhot:2022chi}
J.~Jerhot, B.~D\"obrich, F.~Ertas, F.~Kahlhoefer, T.~Spadaro, {ALPINIST: Axion-Like Particles In Numerous Interactions Simulated and Tabulated}.
\newblock JHEP \textbf{07}, 094 (2022).
\newblock \doi{10.1007/JHEP07(2022)094}.
\newblock {\href{https://arxiv.org/abs/2201.05170}{{arXiv:2201.05170}}} {[hep-ph]}

\bibitem{Ovchynnikov:2023cry}
M.~Ovchynnikov, J.L. Tastet, O.~Mikulenko, K.~Bondarenko, {Sensitivities to feebly interacting particles: public and unified calculations}  (2023).
\newblock {\href{https://arxiv.org/abs/2305.13383}{{arXiv:2305.13383}}} {[hep-ph]}

\bibitem{Curtin:2023skh}
D.~Curtin, J.S. Grewal, {Long Lived Particle Decays in MATHUSLA}  (2023).
\newblock {\href{https://arxiv.org/abs/2308.05860}{{arXiv:2308.05860}}} {[hep-ph]}

\bibitem{Alwall:2006yp}
J.~Alwall, et~al., {A Standard format for Les Houches event files}.
\newblock Comput. Phys. Commun. \textbf{176}, 300--304 (2007).
\newblock \doi{10.1016/j.cpc.2006.11.010}.
\newblock {\href{https://arxiv.org/abs/hep-ph/0609017}{{arXiv:hep-ph/0609017}}}

\bibitem{Dobbs:2001ck}
M.~Dobbs, J.B. Hansen, {The HepMC C++ Monte Carlo event record for High Energy Physics}.
\newblock Comput. Phys. Commun. \textbf{134}, 41--46 (2001).
\newblock \doi{10.1016/S0010-4655(00)00189-2}

\bibitem{Sjostrand:2014zea}
T.~Sj\"ostrand, S.~Ask, J.R. Christiansen, R.~Corke, N.~Desai, P.~Ilten, S.~Mrenna, S.~Prestel, C.O. Rasmussen, P.Z. Skands, {An introduction to PYTHIA 8.2}.
\newblock Comput. Phys. Commun. \textbf{191}, 159--177 (2015).
\newblock \doi{10.1016/j.cpc.2015.01.024}.
\newblock {\href{https://arxiv.org/abs/1410.3012}{{arXiv:1410.3012}}} {[hep-ph]}

\bibitem{MC_PID}
K.~F., S.~Navas, P.~Richardson, T.~Sjöstrand, {43. Monte Carlo Particle Numbering Scheme: \url{https://pdg.lbl.gov/2019/reviews/rpp2019-rev-monte-carlo-numbering.pdf}} (2019)

\bibitem{Bahr:2008pv}
M.~Bahr, et~al., {Herwig++ Physics and Manual}.
\newblock Eur. Phys. J. C \textbf{58}, 639--707 (2008).
\newblock \doi{10.1140/epjc/s10052-008-0798-9}.
\newblock {\href{https://arxiv.org/abs/0803.0883}{{arXiv:0803.0883}}} {[hep-ph]}

\bibitem{Bellm:2019zci}
J.~Bellm, et~al., {Herwig 7.2 release note}.
\newblock Eur. Phys. J. C \textbf{80}(5), 452 (2020).
\newblock \doi{10.1140/epjc/s10052-020-8011-x}.
\newblock {\href{https://arxiv.org/abs/1912.06509}{{arXiv:1912.06509}}} {[hep-ph]}

\bibitem{doxygen}
D.~van Heesch, {Doxygen manal: \url{https://www.doxygen.nl/manual/ }} (2023)

\bibitem{MATHUSLA:2020uve}
C.~Alpigiani, et~al., {An Update to the Letter of Intent for MATHUSLA: Search for Long-Lived Particles at the HL-LHC}  (2020).
\newblock {\href{https://arxiv.org/abs/2009.01693}{{arXiv:2009.01693}}} {[physics.ins-det]}

\bibitem{Gligorov:2017nwh}
V.V. Gligorov, S.~Knapen, M.~Papucci, D.J. Robinson, {Searching for Long-lived Particles: A Compact Detector for Exotics at LHCb}.
\newblock Phys. Rev. D \textbf{97}(1), 015023 (2018).
\newblock \doi{10.1103/PhysRevD.97.015023}.
\newblock {\href{https://arxiv.org/abs/1708.09395}{{arXiv:1708.09395}}} {[hep-ph]}

\bibitem{Gligorov:2018vkc}
V.V. Gligorov, S.~Knapen, B.~Nachman, M.~Papucci, D.J. Robinson, {Leveraging the ALICE/L3 cavern for long-lived particle searches}.
\newblock Phys. Rev. D \textbf{99}(1), 015023 (2019).
\newblock \doi{10.1103/PhysRevD.99.015023}.
\newblock {\href{https://arxiv.org/abs/1810.03636}{{arXiv:1810.03636}}} {[hep-ph]}

\bibitem{Bauer:2019vqk}
M.~Bauer, O.~Brandt, L.~Lee, C.~Ohm, {ANUBIS: Proposal to search for long-lived neutral particles in CERN service shafts}  (2019).
\newblock {\href{https://arxiv.org/abs/1909.13022}{{arXiv:1909.13022}}} {[physics.ins-det]}

\bibitem{ANUBIS_talk_slides}
L.D. Corpe, {Update on (pro)ANUBIS detector proposal: \url{https://indico.cern.ch/event/1216822/contributions/5449255/attachments/2671754/4631593/LCORPE_LLPWorkshop2023_ANUBIS_June2023.pdf }} (2023)

\bibitem{Satterthwaite:2839063}
T.P. Satterthwaite.
\newblock {Sensitivity of the ANUBIS and ATLAS Detectors to Neutral Long-Lived Particles Produced in $pp$ Collisions at the Large Hadron Collider} (2022).
\newblock \urlprefix\url{http://cds.cern.ch/record/2839063}.
\newblock Presented 08 Sep 2022

\bibitem{Pinfold:2019nqj}
J.L. Pinfold, {The MoEDAL Experiment at the LHC\textemdash{}A Progress Report}.
\newblock Universe \textbf{5}(2), 47 (2019).
\newblock \doi{10.3390/universe5020047}

\bibitem{Cerci:2021nlb}
S.~Cerci, et~al., {FACET: A new long-lived particle detector in the very forward region of the CMS experiment}  (2021).
\newblock {\href{https://arxiv.org/abs/2201.00019}{{arXiv:2201.00019}}} {[hep-ex]}

\bibitem{DeVries:2020jbs}
J.~De~Vries, H.K. Dreiner, J.Y. G\"unther, Z.S. Wang, G.~Zhou, {Long-lived Sterile Neutrinos at the LHC in Effective Field Theory}.
\newblock JHEP \textbf{03}, 148 (2021).
\newblock \doi{10.1007/JHEP03(2021)148}.
\newblock {\href{https://arxiv.org/abs/2010.07305}{{arXiv:2010.07305}}} {[hep-ph]}

\bibitem{Dreiner:2020qbi}
H.K. Dreiner, J.Y. G\"unther, Z.S. Wang, {$R$-parity violation and light neutralinos at ANUBIS and MAPP}.
\newblock Phys. Rev. D \textbf{103}(7), 075013 (2021).
\newblock \doi{10.1103/PhysRevD.103.075013}.
\newblock {\href{https://arxiv.org/abs/2008.07539}{{arXiv:2008.07539}}} {[hep-ph]}

\bibitem{Domingo:2022emr}
F.~Domingo, H.K. Dreiner, {Decays of a bino-like particle in the low-mass regime}.
\newblock SciPost Phys. \textbf{14}, 134 (2023).
\newblock \doi{10.21468/SciPostPhys.14.5.134}.
\newblock {\href{https://arxiv.org/abs/2205.08141}{{arXiv:2205.08141}}} {[hep-ph]}

\bibitem{Gunther:2023vmz}
J.Y. G\"unther, J.~de~Vries, H.K. Dreiner, Z.S. Wang, G.~Zhou, {Long-lived neutral fermions at the DUNE near detector}.
\newblock JHEP \textbf{01}, 108 (2024).
\newblock \doi{10.1007/JHEP01(2024)108}.
\newblock {\href{https://arxiv.org/abs/2310.12392}{{arXiv:2310.12392}}} {[hep-ph]}

\bibitem{PIENU:2017wbj}
A.~Aguilar-Arevalo, et~al., {Improved search for heavy neutrinos in the decay $\pi\rightarrow e\nu$}.
\newblock Phys. Rev. D \textbf{97}(7), 072012 (2018).
\newblock \doi{10.1103/PhysRevD.97.072012}.
\newblock {\href{https://arxiv.org/abs/1712.03275}{{arXiv:1712.03275}}} {[hep-ex]}

\bibitem{Bryman:2019bjg}
D.A. Bryman, R.~Shrock, {Constraints on Sterile Neutrinos in the MeV to GeV Mass Range}.
\newblock Phys. Rev. D \textbf{100}, 073011 (2019).
\newblock \doi{10.1103/PhysRevD.100.073011}.
\newblock {\href{https://arxiv.org/abs/1909.11198}{{arXiv:1909.11198}}} {[hep-ph]}

\bibitem{NA62:2020mcv}
E.~Cortina~Gil, et~al., {Search for heavy neutral lepton production in K+ decays to positrons}.
\newblock Phys. Lett. B \textbf{807}, 135599 (2020).
\newblock \doi{10.1016/j.physletb.2020.135599}.
\newblock {\href{https://arxiv.org/abs/2005.09575}{{arXiv:2005.09575}}} {[hep-ex]}

\bibitem{T2K:2019jwa}
K.~Abe, et~al., {Search for heavy neutrinos with the T2K near detector ND280}.
\newblock Phys. Rev. D \textbf{100}(5), 052006 (2019).
\newblock \doi{10.1103/PhysRevD.100.052006}.
\newblock {\href{https://arxiv.org/abs/1902.07598}{{arXiv:1902.07598}}} {[hep-ex]}

\bibitem{Barouki:2022bkt}
R.~Barouki, G.~Marocco, S.~Sarkar, {Blast from the past II: Constraints on heavy neutral leptons from the BEBC WA66 beam dump experiment}.
\newblock SciPost Phys. \textbf{13}, 118 (2022).
\newblock \doi{10.21468/SciPostPhys.13.5.118}.
\newblock {\href{https://arxiv.org/abs/2208.00416}{{arXiv:2208.00416}}} {[hep-ph]}

\bibitem{DELPHI:1996qcc}
P.~Abreu, et~al., {Search for neutral heavy leptons produced in Z decays}.
\newblock Z. Phys. C \textbf{74}, 57--71 (1997).
\newblock \doi{10.1007/s002880050370}.
\newblock [Erratum: Z.Phys.C 75, 580 (1997)]

\bibitem{Sabti:2020yrt}
N.~Sabti, A.~Magalich, A.~Filimonova, {An Extended Analysis of Heavy Neutral Leptons during Big Bang Nucleosynthesis}.
\newblock JCAP \textbf{11}, 056 (2020).
\newblock \doi{10.1088/1475-7516/2020/11/056}.
\newblock {\href{https://arxiv.org/abs/2006.07387}{{arXiv:2006.07387}}} {[hep-ph]}

\bibitem{Canetti:2010aw}
L.~Canetti, M.~Shaposhnikov, {Baryon Asymmetry of the Universe in the NuMSM}.
\newblock JCAP \textbf{09}, 001 (2010).
\newblock \doi{10.1088/1475-7516/2010/09/001}.
\newblock {\href{https://arxiv.org/abs/1006.0133}{{arXiv:1006.0133}}} {[hep-ph]}

\bibitem{Planck:2018vyg}
N.~Aghanim, et~al., {Planck 2018 results. VI. Cosmological parameters}.
\newblock Astron. Astrophys. \textbf{641}, A6 (2020).
\newblock \doi{10.1051/0004-6361/201833910}.
\newblock [Erratum: Astron.Astrophys. 652, C4 (2021)].
\newblock {\href{https://arxiv.org/abs/1807.06209}{{arXiv:1807.06209}}} {[astro-ph.CO]}

\bibitem{FORESEE_github}
F.~Kling, S.~Trojanowski, {FORESEE: \url{https://github.com/KlingFelix/FORESEE}} (2024)

\bibitem{FASER:2018ceo}
A.~Ariga, et~al., {Letter of Intent for FASER: ForwArd Search ExpeRiment at the LHC}  (2018).
\newblock {\href{https://arxiv.org/abs/1811.10243}{{arXiv:1811.10243}}} {[physics.ins-det]}

\bibitem{FASER:2018bac}
A.~Ariga, et~al., {Technical Proposal for FASER: ForwArd Search ExpeRiment at the LHC}  (2018).
\newblock {\href{https://arxiv.org/abs/1812.09139}{{arXiv:1812.09139}}} {[physics.ins-det]}

\bibitem{Ellwanger:2022jtd}
U.~Ellwanger, C.~Hugonie, {Benchmark planes for Higgs-to-Higgs decays in the NMSSM}.
\newblock Eur. Phys. J. C \textbf{82}(5), 406 (2022).
\newblock \doi{10.1140/epjc/s10052-022-10364-3}.
\newblock {\href{https://arxiv.org/abs/2203.05049}{{arXiv:2203.05049}}} {[hep-ph]}

\bibitem{Domingo:2022kfm}
F.~Domingo, S.~Pa\ss{}ehr, {About the bosonic decays of heavy Higgs states in the (N)MSSM}  (2022).
\newblock {\href{https://arxiv.org/abs/2207.05776}{{arXiv:2207.05776}}} {[hep-ph]}

\bibitem{pair_slepton_XS}
A.~Mann, {SUSYCrossSections13TeVslepslep: \url{https://twiki.cern.ch/twiki/bin/view/LHCPhysics/SUSYCrossSections13TeVslepslep}} (2019)

\end{thebibliography}

\end{document}